\documentclass[prl,amsmath,floats,floatfix, twocolumn,
superscriptaddress,nofootinbib,showpacs]{revtex4-1}

\usepackage{amsfonts}
\usepackage{amsmath}
\usepackage{amssymb}
\usepackage{amsthm}
\usepackage{bm}
\usepackage{dcolumn}
\usepackage{epsfig}
\usepackage{graphicx}
\usepackage{graphics}
\usepackage[latin1]{inputenc}
\usepackage{rotating}
\usepackage{comment}
\usepackage{xspace}
\usepackage{subfigure}

\usepackage[dvipsnames, usenames]{xcolor}
\definecolor{linkcolor}{rgb}{0.0,0.3,0.5}
\usepackage[hypertexnames=false, unicode, colorlinks=true, linkcolor=linkcolor,
citecolor=linkcolor, filecolor=linkcolor,urlcolor=linkcolor,
pdfusetitle]{hyperref}

\newcommand{\model}{\texttt{BHPTNRSur2dq1e3}\xspace}
\newcommand{\modelv}{\texttt{BHPTNRSur1dq1e4}\xspace}

\usepackage[normalem]{ulem} 

\begin{document}

\title{Gravitational wave surrogate model for spinning, intermediate mass ratio binaries based on perturbation theory and numerical relativity}

\author{Katie Rink} \affiliation{Department of Physics and Center for Scientific Computing \& Visualization Research, University of Massachusetts, Dartmouth, MA 02747} \affiliation{Department of Physics and Weinberg Institute for Theoretical Physics, University of Texas at Austin, TX 78712}

\author{Ritesh Bachhar} \affiliation{Department of Physics and Center for Computational Research, East Hall, University of Rhode Island, Kingston, RI 02881}

\author{Tousif Islam} \affiliation{Department of Physics and Center for Scientific Computing \& Visualization Research, University of Massachusetts, Dartmouth, MA 02747}
\affiliation{Kavli Institute for Theoretical Physics,\\University of California Santa Barbara, Kohn Hall, Lagoon Rd, Santa Barbara, CA 93106}
\affiliation{Kavli Institute for Theoretical Physics, University of California Santa Barbara, CA 93106}

\author{Nur E. M. Rifat} \affiliation{Department of Physics and Center for Scientific Computing \& Visualization Research, University of Massachusetts, Dartmouth, MA 02747}
\affiliation{Department of Physics, University of Virginia, 1827 University Avenue, Charlottesville, VA 22903}

\author{Kevin Gonz\'alez-Quesada} \affiliation{Department of Physics and Center for Scientific Computing \& Visualization Research, University of Massachusetts, Dartmouth, MA 02747}
\affiliation{Department of Physics and Center for Education and Research in Cosmology and Astrophysics (CERCA), Case Western Reserve University, Cleveland, OH 44106}

\author{Scott E. Field} \affiliation{Department of Mathematics and Center for Scientific Computing \& Visualization Research, University of Massachusetts, Dartmouth, MA 02747}

\author{Gaurav Khanna} \affiliation{Department of Physics and Center for Computational Research, East Hall, University of Rhode Island, Kingston, RI 02881}
\affiliation{Department of Physics and Center for Scientific Computing \& Visualization Research, University of Massachusetts, Dartmouth, MA 02747}

\author{Scott A. Hughes}
\affiliation{Department of Physics, Massachusetts Institute of Technology, Cambridge, MA 02139} 

\author{Vijay Varma}
\affiliation{Department of Mathematics and Center for Scientific Computing \& Visualization Research, University of Massachusetts, Dartmouth, MA 02747}

\begin{abstract}
We present \model, a reduced order surrogate model of gravitational waves emitted from binary black hole (BBH) systems in the comparable to large mass ratio regime with aligned spin ($\chi_1$) on the heavier mass ($m_1$). We trained this model on waveform data generated from point particle black hole perturbation theory (ppBHPT) with mass ratios varying from $3 \leq q \leq 1000$ and spins from $-0.8 \leq \chi_1  \leq 0.8$. The waveforms are 13,500 $m_1$ long and include all $\ell \leq 4$ spin-weighted spherical harmonic modes except the $(4,1)$ and $m=0$  modes. We find that for binaries with $\chi_1\lesssim-0.5$, retrograde quasi-normal modes are significantly excited, thereby complicating the modeling process. To overcome this issue, we introduce a domain decomposition approach to model the inspiral and merger-ringdown portion of the signal separately. The resulting model can faithfully reproduce ppBHPT waveforms with a median time-domain mismatch error of $8 \times 10^{-5}$. We then calibrate our model with numerical relativity (NR) data in the comparable mass regime ($3 \leq q \leq 10$). By comparing with spin-aligned BBH NR simulations at $q=15$, we find that the dominant quadrupolar (subdominant) modes agree to better than $\approx 10^{-3}$ ($\approx 10^{-2}$) when using a time-domain mismatch error, where the largest source of calibration error comes from the transition-to-plunge and ringdown approximations of perturbation theory. Mismatch errors are below $\approx 10^{-2}$ for systems with mass ratios between $6 \leq q \leq 15$ and typically get smaller at larger mass ratio. Our two models -- both the ppBHPT waveform model and the NR-calibrated ppBHPT model -- will be publicly available through \texttt{gwsurrogate} and the Black Hole Perturbation Toolkit packages.
\end{abstract}

\maketitle

\section{Introduction}
\label{Sec:Introduction}

To date, the Advanced Laser Interferometer Gravitational-Wave Observatory (LIGO) \cite{LIGOScientific:2014pky},  Advanced Virgo \cite{VIRGO:2014yos}, and KAGRA \cite{KAGRA:2020tym} Collaboration (LVK) have detected 90 compact binary coalescence events \cite{GWTC12019, LIGOScientific:2020ibl, LIGOScientific:2021usb, KAGRA:2021vkt}. These systems are mostly in the comparable mass ratio regime, yet a handful of events (such as $ \mathrm{GW190814, \ GW200210\_092254, \ and \ GW191219\_163120} $) have measured mass ratios on the order $q \sim 10$ \cite{LIGOScientific:2020zkf, KAGRA:2021vkt}, where $q = m_1 / m_2 \geq 1$, $m_1$ refers to the primary black hole's mass, and $m_2$ the secondary black hole's mass. The posterior distributions computed for some events extend to even larger $q$ values outside the NR calibration range of current waveform models~\cite{KAGRA:2021vkt}.  To reduce systematic bias when analyzing the current set of gravitational wave (GW) observations, as well as prepare for the multitude of large mass ratio events we expect to observe with next-generation detectors \cite{LISAConsortiumWaveformWorkingGroup:2023arg, Reitze:2019iox, Punturo:2010zz}, we must extend our waveform models to accurately span several orders of magnitude in mass ratio. Moreover, the improved sensitivities of future detectors will place increasing demands on waveform model accuracy standards. It is necessary we develop extensive waveform models across the mass ratio parameter space that are indistinguishable from these high signal-to-noise-ratio (SNR) signals for future matched-filtering searches and parameter estimation \cite{Purrer:2019jcp, Ferguson_2021}.

The most accurate waveforms come from NR simulations \cite{ferguson2023second, Mroue:2013xna, Boyle_2019, L_ffler_2012, Jani:2016wkt, Healy:2017psd, Healy:2019jyf, Healy:2020vre, Healy:2022wdn, neilsen2022challenges}. However, these simulations are computationally expensive and only cover a small portion of the needed parameter space. Currently, NR waveforms with sufficient accuracy and durations are mostly limited to $q < 10$. Some recent breakthroughs, however, have made it possible to perform  NR simulations for high mass ratio inspirals, including up to mass ratios of about $30$~\cite{Yoo:2022erv,Giesler:2022inPrep,Lousto:2020tnb}. A major source of GWs we expect to observe with next-generation detectors, especially space-borne detectors such as the Laser Interferometer Space Antenna (LISA), are intermediate-to-extreme mass ratio inspirals (I-EMRIs). These systems are typically composed of a stellar-to-intermediate sized black hole orbiting around an intermediate-to-supermassive black hole ($q \gtrsim 100$) with inspirals lasting for up to thousands of orbits in the detector's sensitivity band. Even if NR codes could overcome the prohibitively expensive computational cost of modeling such systems, they lose accuracy as mass ratio increases \cite{Purrer:2019jcp}. 

Motivated by the growing evidence for the ubiquity of nonzero spins in the GWTC-3 catalog \cite{KAGRA:2021vkt}, it is essential we include spin in waveform models to get the most accurate science out of our post-detection analyses. The diverse parameter-space of I-EMRI systems we expect to observe with next-generation detectors will offer a rich new dataset to probe many open astrophysical questions, such as: how black holes form in the pair-instability mass gap \cite{Gair:2010dx}, what formation channels lead to supermassive black hole binaries \cite{Bellovary:2019nib}, or how environmental effects from accretion disks impact the morphology of a GW signal \cite{Amaro-Seoane:2007osp}. As our detection horizon expands to higher and higher redshifts, we may be able to detect dynamical mergers within the accretion disks of active galactic nuclei (AGN). There may already be evidence of such dynamics from X-ray observations of the outflowing material around an AGN which exhibits quasi-periodic changes in brightness, indicative of an intermediate mass black hole on an eccentric and inclined orbit periodically passing through the disk \cite{pasham2024case}. I-EMRI GW detections, along with any electromagnetic counterparts or precursors, will contribute tremendously to probing this population of intermediate-mass-black-holes and informing studies on stellar evolution in the upper end of the BH mass-gap \cite{Mehta_2022, McKernan:2019beu}. Moreover, the long inspirals of I-EMRI signals will give us a detailed map of the spacetime around massive BHs, which we can use to test alternative theories of gravity \cite{LISAConsortiumWaveformWorkingGroup:2023arg}.

Much progress has been made in finding accurate approximate modeling methods to generate waveforms needed before these next-generation detectors can observe and accurately estimate the parameters of these signals. Approximations such as post-Newtonian expansion \cite{Lorentz1937} (see \cite{Blanchet:2013haa} for a more recent review) and point-particle perturbation theory \cite{Teukolsky:1972PhysRevLett.29.1114} have proven promising in the slow motion and high mass ratio regime, respectively, while effective-one-body \cite{Buonanno:1999PhysRevD.59.084006, Bohe:2016gbl, Cotesta:2018fcv, Cotesta:2020qhw, Pan:2013rra, Babak:2016tgq, Ossokine:2020kjp, Damour:2014sva, Nagar:2019wds, Nagar:2020pcj, Riemenschneider:2021ppj, Khalil:2023kep, Pompili:2023tna, Ramos-Buades:2023ehm, vandeMeent:2023ols} and phenomenological methods \cite{Ajith_2007, Husa:2015iqa, Khan:2015jqa, London:2017bcn, Khan:2018fmp, Hannam:2013oca, Khan:2019kot, Pratten:2020ceb, Estelles:2020osj, Estelles:2020twz, Estelles:2021gvs, Hamilton:2021pkf} can cover the entire waveform regime. However, the accuracy of these models is limited by approximations made within their frameworks.

More recently, surrogate modeling efforts have exploited data-driven techniques to build waveform models across parameter space from a set of NR training waveforms ~\cite{Field:2013cfa, Blackman:2015pia, Blackman:2017dfb, Blackman:2017pcm, varma2019surrogate2, varma2019surrogate}. This method allows us to construct a fast waveform model from a smaller set of computationally expensive simulations without sacrificing accuracy~\cite{Field:2013cfa}. Furthermore, these surrogate models are valid throughout the entire regime covered by the waveforms used for their training data. This makes surrogate modeling an especially promising avenue for building I-EMRI waveform models, as there, too, the underlying simulation data from point particle black hole perturbation theory (ppBHPT) requires computationally expensive numerical solvers.  Rifat \textit{et al.}~\cite{Rifat:2019ltp} first demonstrated this by building perturbation theory waveforms up to $q = 10^4$ and calibrating to NR waveforms in the comparable mass ratio regime. This model was then updated by Islam \textit{et al.} \cite{1d_updated} using an improved transition-to-plunge model as well as improving the NR calibration step to include higher harmonic modes. Here, we continue the development of this modeling effort by extending our previous work, notably the nonspinning model \modelv, to include spin on the primary black hole. 

In this paper, we extend \modelv to cover a two-dimensional parameter space of non-precessing binary black hole systems (BBHs) with a spinning primary BH and a non-spinning secondary BH. The model is developed for a wide range of mass ratios varying from $q=3$ to $q=1000$, along with a spinning primary BH $\chi_1 \in [-0.8, 0.8]$. Our model is trained on waveform data generated by solving the Teukolsky equation~\cite{sundararajan2007towards, sundararajan2008towards, sundararajan2010binary, zenginouglu2011null}, representing all three inspiral, merger, and ringdown phases of the BBH evolution. Each waveform is 13,500$m_1$ in duration,  corresponding to between 56 and 425 orbital cycles depending on the specific values of $(q,\chi_1)$ being evaluated. We model the GW strain, $h_{\ell m}(t;q,\chi_1)$ for the spin-weighted spherical harmonic modes $(\ell, m) = \{(2, \{2,1\}), (3,\{3,2,1\}), (4,\{4,3,2\})\}$ and their $m<0$ counterparts. By calibrating our ppBHPT waveform model to NR in the comparable mass ratio regime while including the exact $q \rightarrow \infty$ perturbation theory result, we achieve an accurate waveform model for intermediate mass ratio systems. 

The rest of this paper is organized as follows. We begin by discussing our training waveforms and the Teukolsky equation solver used to generate them, detailing modeling challenges in Section \ref{Sec:training_data}. In Section \ref{Sec:buildSurrogate} we outline the surrogate-building process with a detailed discussion of the accuracy of the model. Section \ref{sec:calibration} discusses how we calibrate our surrogate waveforms to those generated using NR in the comparable-to-intermediate mass ratio regime. Finally, in Section \ref{sec:summary} we conclude with a summary of our work as well as future endeavors. The \model model~\footnote{In this paper, we build a model for waveforms computed by numerically solving the Teukolsky equation and a second (closely related) model that introduces amplitude and phase calibration parameters that are set by matching to NR. We shall refer to both models as \model when it is clear from context.} will be made publicly available as part of both the Black Hole Perturbation Toolkit~\cite{BHPToolkit} and GWSurrogate~\cite{gwsurrogate}.

\begin{figure}[h!]
    \centering
    \includegraphics[width=\columnwidth]{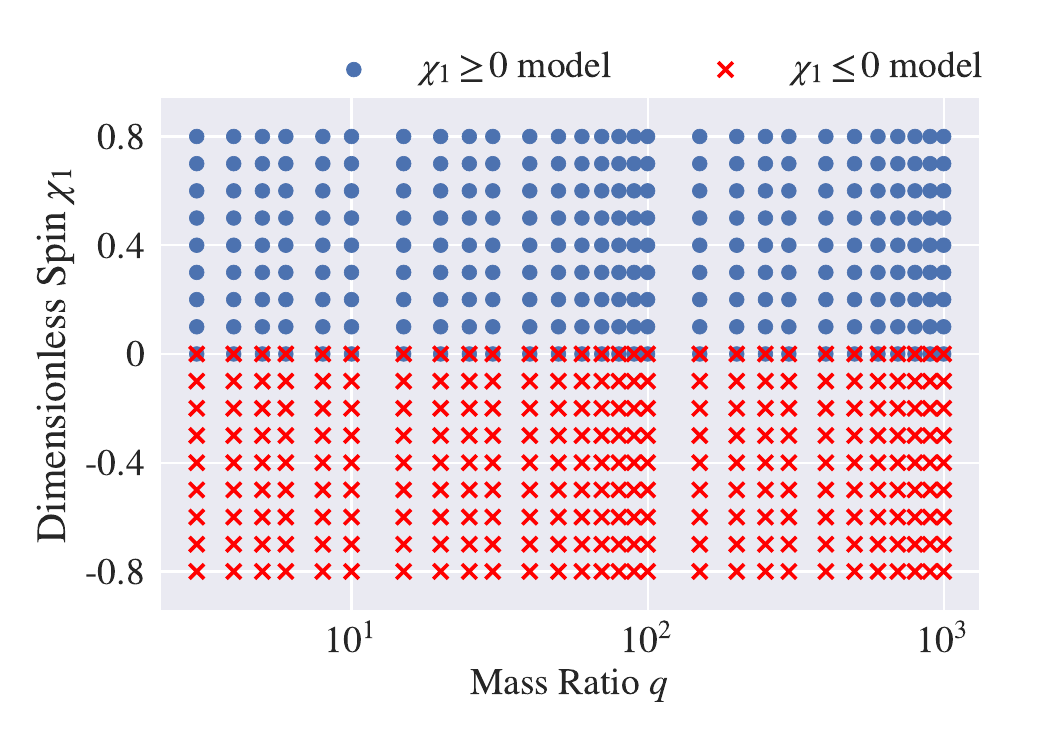}
    \caption[The parameter space sampled by our set of waveform training data.]{The parameter space sampled by our waveform training data. Each marker represents an individual ppBHPT waveform of a given $q$ and $\chi_1$ in our initial data set. For reasons discussed in Sec.~\ref{sec:retrograde_modes}, the positive and negative spin regions are modeled separately, overlapping at $\chi_1 = 0$.}
    \label{fig:training_bank}
\end{figure}

\section{Training Waveforms Using Perturbation Theory}
\label{Sec:training_data}

We generate the surrogate model's training data using ppBHPT. In this approach, the smaller, nonspinning, black hole of mass $m_{2}$ is modeled as a point-particle with no internal structure, moving in the space-time of the larger Kerr black hole of mass $m_{1}$ and dimensionless spin $\chi_1$. In the large mass ratio limit ($q \gg 1$), the system's dynamics are well described using Kerr black hole perturbation theory.  

We implemented this ppBHPT approach in two steps. First, we compute the trajectory taken by the point particle and then use that trajectory to compute the GW emission. The models, methods, and numerics used to generate our training data are essentially identical to those used for the previously built nonspinning model \modelv. Here, we provide a brief summary and refer to Ref.~\cite{1d_updated} for additional details.

The particle's motion can be characterized by three distinct regimes -- an initial adiabatic inspiral, a late-stage geodesic plunge into the horizon, and a transition regime between those two~\cite{ori2000transition,Hughes:2019zmt,sundararajan2010binary,Apte_2019}. In the initial adiabatic inspiral, the particle follows a sequence of geodesic orbits, driven by radiative energy and angular momentum losses computed by solving the frequency-domain Teukolsky equation~\cite{10.1143/PTP.112.415,10.1143/PTP.113.1165,10.1143/PTP.95.1079,ThorweThesis} with the open-source code GremlinEq~\cite{gremlin,OSullivan:2014ywd,Drasco_2006} from the Black Hole Perturbation Toolkit~\cite{BHPToolkit}. It should be noted that our inspiral model does not include the effects of the conservative or second-order self-force~\cite{Hinderer_2008}, although once these post-adiabatic corrections are known (see, for example, Refs.~\cite{Gralla_2012,Pound_2012,Pound:2019lzj,wardell2023gravitational}) they could be easily incorporated to improve the accuracy of the inspiral's phase. The inspiral trajectory is then extended to include a plunge geodesic and a smooth transition region following the generalized Ori-Thorne procedure \cite{Apte_2019} (hereafter, the ``GOT'' algorithm). Crucial for our purpose, GOT introduces a correction that smooths a discontinuity in the evolution of an inspiral's integrals of motion as presented in the original Ori-Thorne model~\cite{ori2000transition} (see Sec.IV A 2 of Ref. \cite{Apte_2019}). Note that we use ``Model 2'' from Ref. \cite{Apte_2019} for this smoothing.

With the trajectory of the perturbing compact body fully specified, we then solve the inhomogeneous Teukolsky equation in the time domain while feeding the trajectory information into the particle source term of the equation. Details regarding the formulation of the Teukolsky equation and its numerical discretization with a finite-difference numerical evolution scheme can be found in our earlier work~\cite{sundararajan2007towards, sundararajan2008towards, sundararajan2010binary, zenginouglu2011null,mckennon2012high}. The gravitational waveform, represented as a 
complex strain $h = h_{+} - i h_{\times}$,
\begin{equation}
    h(t, \theta, \phi; q, \chi_1) = \sum^{\infty}_{\ell=2} \sum_{m=-\ell}^{\ell}
        h_{\ell m}(t; q, \chi_1) ~_{-2}Y_{\ell m}(\theta, \phi),
\label{eq:spherical_harm}
\end{equation}
is computed by the Teukolsky solver. Here, $_{-2}Y_{\ell m}$ is the spin-weight $-2$ spherical harmonic, $h_+$ ($h_{\times}$) is the plus (cross) polarization of the waveform, and $\theta$ and $\phi$ are the polar and azimuthal angles. Due to the BBH system's orbital-plane symmetry, the $m<0$ modes can be readily obtained from the complex-conjugated $m>0$ modes as $h_{\ell, -m} = (-1)^{\ell} h_{\ell m}^*$. Unless otherwise specified, all quantities with dimension appearing in the ppBHPT framework masses are given in units of the background Kerr black hole $m_1$, which is taken to be the mass scale.

We use our Teukolsky solver to generate training waveforms sampling the parameter space $3 \leq q \leq 10^3$ and $| \chi_1 | \leq 0.8$. The values of $(q,\chi_1)$ were chosen on a logarithmic scale for $q$ and uniformly over the range of spins. A total of $476$ ppBHPT waveforms were computed across the parameter space, and the complete set of training waveforms can be seen in Figure \ref{fig:training_bank}.

For mass ratios $q \lesssim 20$, the GOT algorithm results in a small jump in the point-particle's velocity as it exits the adiabatic inspiral and begins to plunge. This jump results in a small, non-physical oscillation in the waveform's amplitude and phase, especially in some of the higher-order modes, like those shown in Fig. 1 in Ref.~\cite{1d_updated}. We employ the smoothing procedure described in paper \cite{1d_updated} to remove these spurious artifacts.

\section{Surrogate model for ppBHPT waveforms}
\label{Sec:buildSurrogate}

This section describes our method to build a surrogate model of ppBHPT waveforms starting from the 476 training waveforms described in Sec.~\ref{Sec:training_data}. Our model is constructed using a combination of methodologies proposed in Refs.~\cite{Blackman:2015pia,Field:2013cfa,Purrer:2014fza,Varma:2018mmi}. In particular, our methodology follows the same steps used to build the nonspinning model \modelv \cite{1d_updated}, with three notable differences. First, our parametric fits are performed using Gaussian Process Regression (also used in spin-aligned NR surrogate models~\cite{Varma:2018mmi} and remnant surrogate models~\cite{Varma:2018aht, varma2019surrogate}) instead of smoothing splines. Second, for $\chi_1 \lesssim -0.6$, we observe significant excitation of retrograde modes in the ringdown signal, necessitating a temporal subdomain partitioning strategy that, to our knowledge, has not appeared in the GW modeling literature. Finally, also to ameliorate modeling difficulties due to retrograde modes, we found more accurate models could be achieved by building two separate models for positive and negative spins following a parametric domain decomposition strategy inspired by earlier works~\cite{Gadre:2022sed,smith2016fast}.

\subsection{Data decomposition and alignment}
\label{subsec:dataAlignement}

After obtaining the ppBHPT waveform training data by the methods summarized in Sec.~\ref{Sec:training_data}, we perform the following steps before building the surrogate model. We first find the time for each waveform at which the quadrature sum of all modeled modes,
\begin{equation}
\label{eq:amplitude}
    A_{\rm tot}(\tau) = \sqrt{\sum_{\ell m}|h_{\ell m}(\tau)|^2} \,,
\end{equation}
is maximized. Here, the sum is taken over all modeled modes instead of all available modes from the ppBHPT simulation. The peak, $\tau_{\rm peak}$, is found by fitting a quadratic function to the nearest four points (two on both sides) of the peak found on the discrete time grid. We transform the time axis, $t=\tau - \tau_{\rm peak}$, so that the maximum of the total amplitude occurs at $t=0$ for all waveforms. Using a cubic spline, each waveform harmonic mode is interpolated onto a common time window from $-13500 m_1$ to $125 m_1$ with a uniform spacing of $\Delta t = 0.1 m_1$.  After performing this temporal alignment, we rotate each waveform about $\hat{z}$ (taken to be perpendicular to the orbital plane) such that at $t = -13500 m_1$, the phase of the $(2,2)$ mode for all waveforms is set to $0$. This choice doesn't uniquely define the rotational freedom, which is fixed by imposing the phase of the $(2,1)$ mode to lie between [$0, \pi$). 

One final step before the surrogate-building process is to transform the rapidly varying complex gravitational wave strain as a function of time into slowly varying functions for ease of model build. Therefore, we decompose each mode's strain 
\begin{equation}
    h_{\ell m}(t ; q, \chi_1)=A_{\ell m}(t ; q, \chi_1) \exp \left(-\rm{i} \phi_{\ell m}(t ; q, \chi_1)\right) \,,
\end{equation}
into an amplitude, $A_{\ell m}(t; q, \chi_1)$, and phase, $\phi_{\ell m}(t; q, \chi_1)$. The full set of waveform {\em data pieces} we model are $A_{\ell m}(t; q, \chi_1)$ and $\phi_{\ell m}(t; q, \chi_1)$. This choice differs from the previous model~\cite{1d_updated} that defines the data pieces in the co-orbital frame. We did not need to pursue this more complicated choice here as the resulting model (cf. Fig.~\ref{fig:loocv_full}) is accurate. Furthermore, our phase data pieces are well defined throughout the parameter space as we do not model equal mass systems, where $\phi_{21}$ can become undefined.

\begin{figure}[h]
    \centering
    \includegraphics[width=\columnwidth]   
    {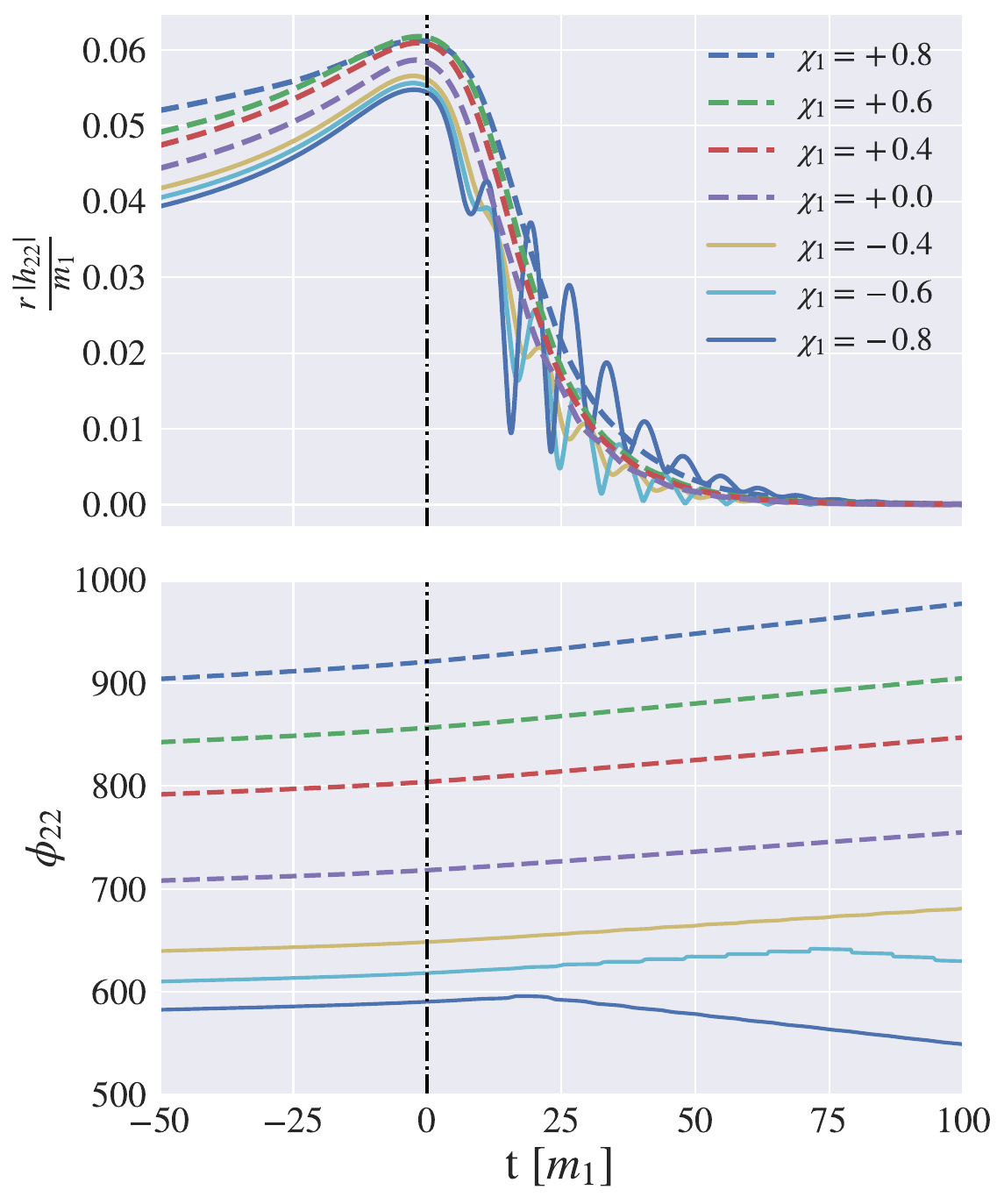}
    \caption[The $(2,2)$ mode late-time phase evolution for a $q = 25$ system.]{The $(2,2)$ mode late-time amplitude (upper panel) and phase (lower panel) evolution for a $q = 25$ system. Here $t = 0$ marks the peak of the waveform (demarcated by a vertical dash-dot line). Each line represents a different spin case.  Starting around $t = 20 m_1$, qualitatively distinct features appear in the ringdown signal for $\chi_1 \lesssim -0.6$, which we attribute to excited retrograde quasi-normal modes in the perturbation theory computation. This poses a significant modeling challenge. Appendix~\ref{app:retrograde} discusses the appearance of retrograde modes in greater detail. }
    \label{fig:ringdown}
\end{figure}

\subsection{Domain decomposition techniques for modeling retrograde ringdown modes}
\label{sec:retrograde_modes}

\subsubsection{Appearance of retrograde modes}

Figure \ref{fig:ringdown} offers us an understanding of how our training data behaves as we vary the spin at a fixed mass ratio. The data has already been aligned according to the procedure described in Sec.~\ref{subsec:dataAlignement}. As the figure shows, one of the main challenges will be accurately modeling the ringdown. Figure \ref{fig:ringdown} shows the $(2,2)$ mode's amplitude and phase evolution of a $q=25$ system for several $\chi_1$ values. Starting around $t = 20 m_1$, qualitatively distinct features appear in the ringdown signal for $\chi_1 \lesssim -0.6$. These features are due to exciting retrograde quasi-normal modes in addition to the usual prograde mode~\cite{taracchini2014small,lim2019exciting}. Appendix~\ref{app:retrograde} considers the physicality of this feature in more detail.

Despite significant experimentation, we could not build accurate surrogate models when attempting a global parametric fit over the full parameter space when including ringdown signals with retrograde modes. To achieve accurate models, we have implemented two domain decomposition strategies described below. When combined together, the resulting domain decomposition will quarantine off the ringdown signal for $\chi_1 <0$ systems, allowing for the overall model to be much more accurate as compared to building a single model.

\subsubsection{Parameter space domain decomposition}

In the domain decomposition approach, we break up the large parameter space into two smaller subdomains defined by $\chi_1 \geq 0$ and $\chi_1 \leq 0$. The resulting subdomains are shown in Fig.~\ref{eq:spherical_harm}. We build two distinct models on each patch, and the positive spin sub-model is used when evaluating at  $\chi_1=0$ as it provides slightly better accuracy.

\subsubsection{Temporal domain decomposition}
\label{subseb:temporal_domains}

Further improvements in ringdown modeling can be achieved by a temporal domain decomposition that separates the inspiral regime of the waveform from the ringdown one. Here we partition the time domain as $[-13500,0] \, \bigcup \, [0,125]$. Training data on each subdomain is simply the aligned waveform data restricted to its subdomain. We implement this approach by using two masking functions, $H(t)$ and its complement $1 - H(t)$, where $H(t)$ is the Heaviside step function. This effectively doubles the number of training waveforms as we train on $\{H(t) h_{\ell m}\}$ and $\{(1-H(t)) h_{\ell m}\}$ separately. 
Our approach yields at least two orders of magnitude better results in the case of $\chi_1 \leq 0$ than those obtained from a surrogate built without temporal domain decomposition. For scenarios where $\chi_1 \geq 0$, we also observe modest improvement in modeling accuracy using the domain decomposition method.

\subsection{Building the surrogate model}

For each $\chi_1 \leq 0$ and $\chi_1 \geq 0$ submodel, we construct a temporal empirical interpolant (EI), a greedy algorithm that picks the most representative basis functions and time nodes~\cite{Maday:2009,chaturantabut2010nonlinear, Field:2013cfa, Canizares:2014fya}. 
The empirical interpolant gives a compact representation for each data piece (and hence the full waveform) in the training set by permitting the full time-series to be reconstructed through a significantly sparser sampling defined by the EI nodes. We choose the number of basis functions (equivalently, the number of EI nodes) for each amplitude and phase data piece
to faithfully represent the data as assessed by a leave-one-out cross-validation study. This results in about 10 to 20 basis functions for each data piece.
We also visually inspect each basis vector to ensure they are free from noise. Following Ref.~\cite{1d_updated}, we put no restriction on the location of EI nodes. We note that due to the temporal decomposition discussed in Sec.~\ref{subseb:temporal_domains}, EI nodes corresponding to times before and after $t=0$ are associated with different temporal domains, and so effectively separate models.

\begin{figure}[h]
    \centering
    \includegraphics[width=8.5cm]
    {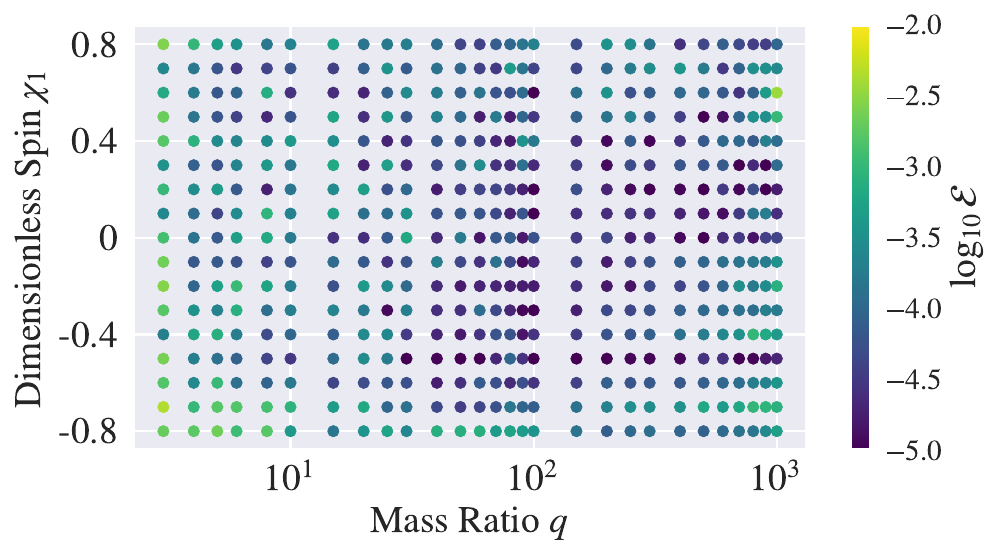}
    \caption{Leave-one-out cross-validation (out-of-sample) errors for our surrogate model
    of ppBHPT waveforms using all available modeled modes. Each dot represents the time-domain mismatch error, ${\cal E}$, computed from Eq.~\eqref{eq:L2} when a training waveform of the corresponding $(q, \chi_1)$ was excluded from the surrogate model building process. Results are shown across the entire parameter space, including the parameter domain boundary points.}
    \label{fig:loocv_full}
\end{figure}

The final surrogate-building step is to construct parametric fits for each data piece at each EI node over the two-dimensional parameter space after performing a logarithmic transformation of $q$~\cite{Varma:2018aht, Varma:2018mmi}. While previous ppBHPT models used smoothing spline fits~\cite{Rifat:2019ltp,1d_updated}, we found splines did not sufficiently mitigate overfitting problems in the presence of noise for our spinning model. We therefore chose to use a Gaussian process regression (GPR) method \cite{GPRbook}, following the implementation and settings of Ref. \cite{Varma:2018aht}.

\begin{figure*}
    \centering
    \includegraphics[width=.99\textwidth]{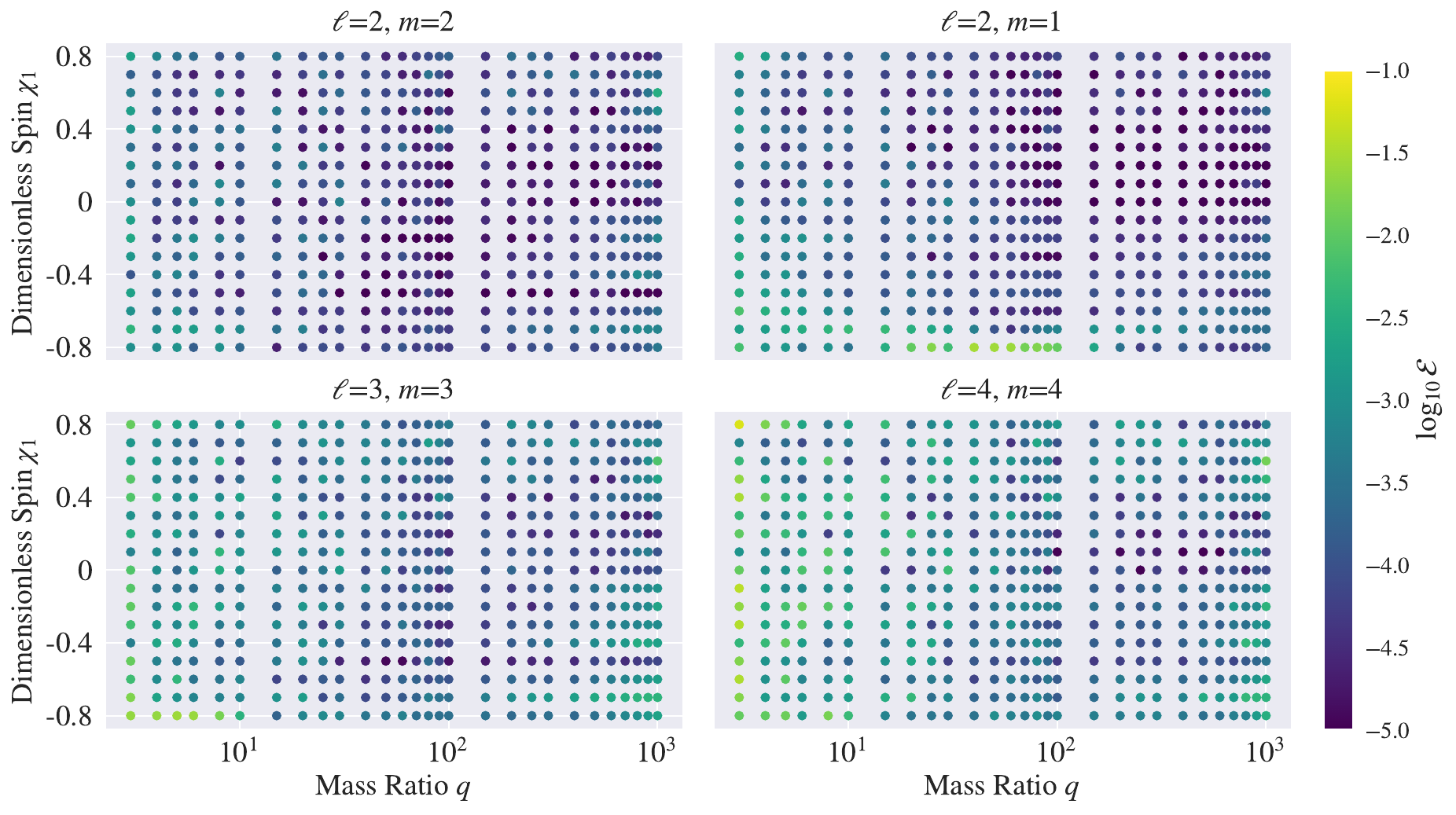}
    \caption{Leave-one-out cross-validation (out-of-sample) error for our surrogate of ppBHPT waveforms assessed for the individual modes $(2,2)$, $(2,1)$, $(3,3)$, and $(4,4)$.  Each dot represents the time-domain mismatch error, ${\cal E}$, computed from Eq.~\eqref{eq:L2}, when a training waveform of the corresponding $(q, \chi_1)$ was excluded from the surrogate model building process. Results are shown across the entire parameter space, including the parameter domain boundary points.}
\label{fig:loocv_modes}
\end{figure*}

Our model, \model, takes two parameters, namely mass ratio $q$ and spin $\chi_1$ as input. We evaluate the GPR fits at the given point in the parameter space, thereby obtaining waveform datapiece evaluations at each EI node. We then use the empirical interpolant representation to efficiently evaluate the amplitude and phase data pieces onto the target time grid, combining them using $h^{\rm S}_{\ell m} = A^{\rm S}_{\ell m}{\rm exp}(-i \phi^{\rm S}_{\ell m})$. The full surrogate model of the gravitational strain in the source frame is written as,
\begin{equation}
    h^{\rm S}(t, \theta, \phi; q, \chi_1) = \sum^{\infty}_{\ell=2} \sum_{m=-\ell}^{\ell}
        h^{\rm S}_{\ell m}(t; q, \chi_1) ~_{-2}Y_{\ell m}(\theta, \phi),
\end{equation}
where $h^{\rm S}(t, \theta, \phi; q, \chi_1)$ represents the full surrogate prediction of the gravitational wave.

\subsection{Surrogate model errors}
\label{sec:results_ppBHPT}

In this section, we investigate the accuracy of our time domain surrogate model of point-particle black hole perturbation theory waveforms.

Assume $h_1(t)$ and $h_2(t)$ represent two complex gravitational-wave strain signals in the time-domain that have already been aligned in time and phase by the process described in Sec~\ref{subsec:dataAlignement}. When comparing $h_1$ to $h_2$, and viewing $h_2$ as the approximation to $h_1$, we compute
\begin{equation}\label{eq:L2}
    \mathcal{E}[h_{1}, h_2] = 
    \frac{1}{2}
    \frac{\sum_{\ell m} \int_{t_{\rm min}}^{t^{\rm max}} |\delta h_{\ell m}(t)|^{2}dt}{ \sum_{\ell m} \int_{t_{\rm min}}^{t^{\rm max}} | h_{1,\ell m}(t)|^2 dt } \,,
\end{equation}
where $\delta h_{\ell m}(t) = h_{1,\ell m}(t) - h_{2,\ell m}(t)$ measures the contribution to the error from each mode. The quantity $\mathcal{E}[h_{1}, h_2]$  is related to the weighted average of the white-noise mismatch error over the sky \cite{Blackman:2017dfb}.

\begin{figure*}
    \centering
    \includegraphics[width=.99\textwidth]{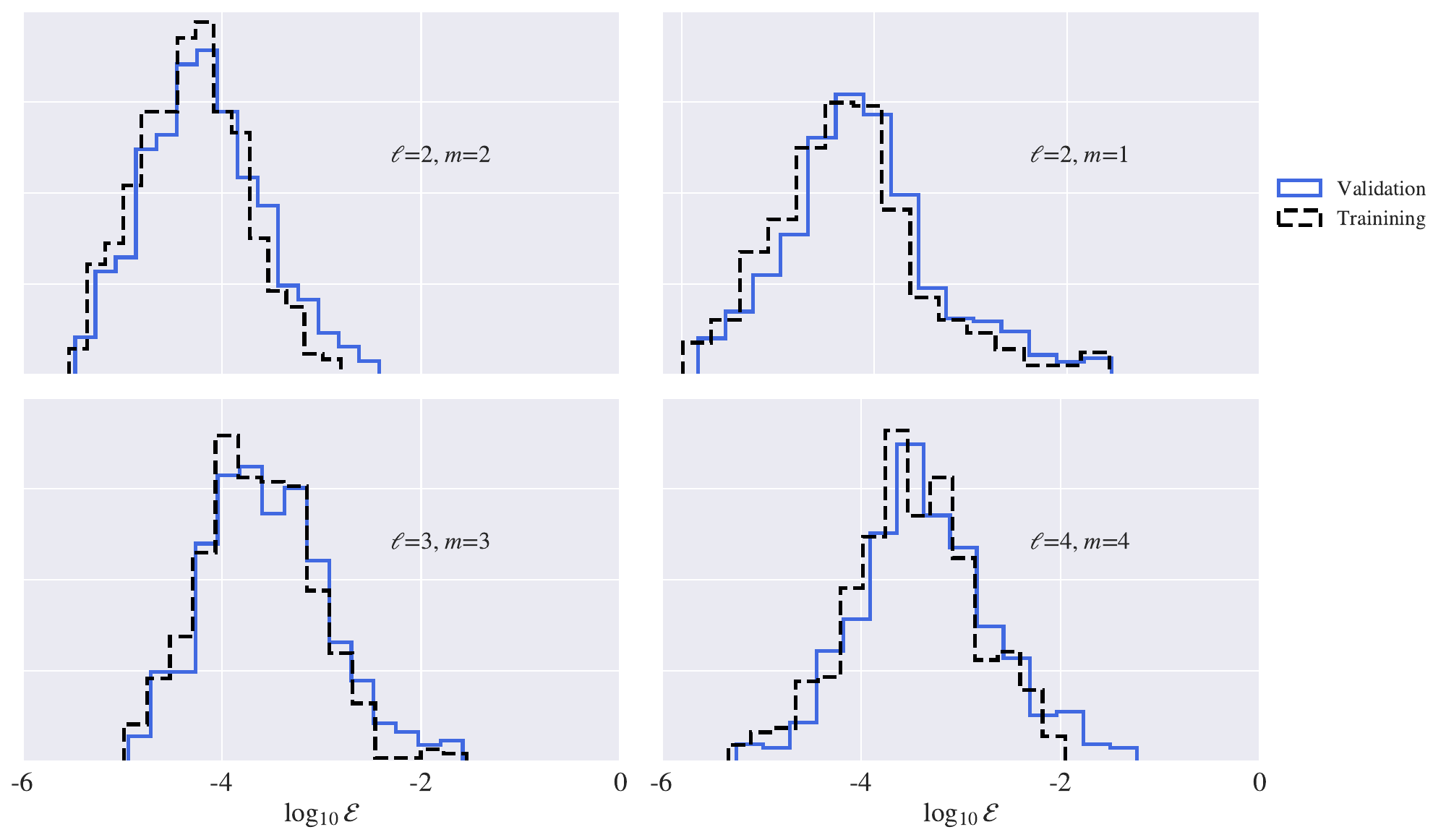}
    \caption{Errors in our ppBHPT surrogate waveform model assessed for the individual modes $(2,2)$, $(2,1)$, $(3,3)$, and $(4,4)$. Error histogram of time-domain mismatch, $\mathcal{E}$, computed from Eq.~\eqref{eq:L2} for both the training (represented by black-dashed) and validation errors (depicted in blue-solid). Histograms are normalized such that the area under the curve is 1. We note that errors for parameter domain boundary points are included in the plot even though the model is effectively extrapolating when validation errors are computed.}
\label{fig:loocv_vs_training}
\end{figure*}

\begin{figure}[h!]
    \centering
    \includegraphics[width=8cm]{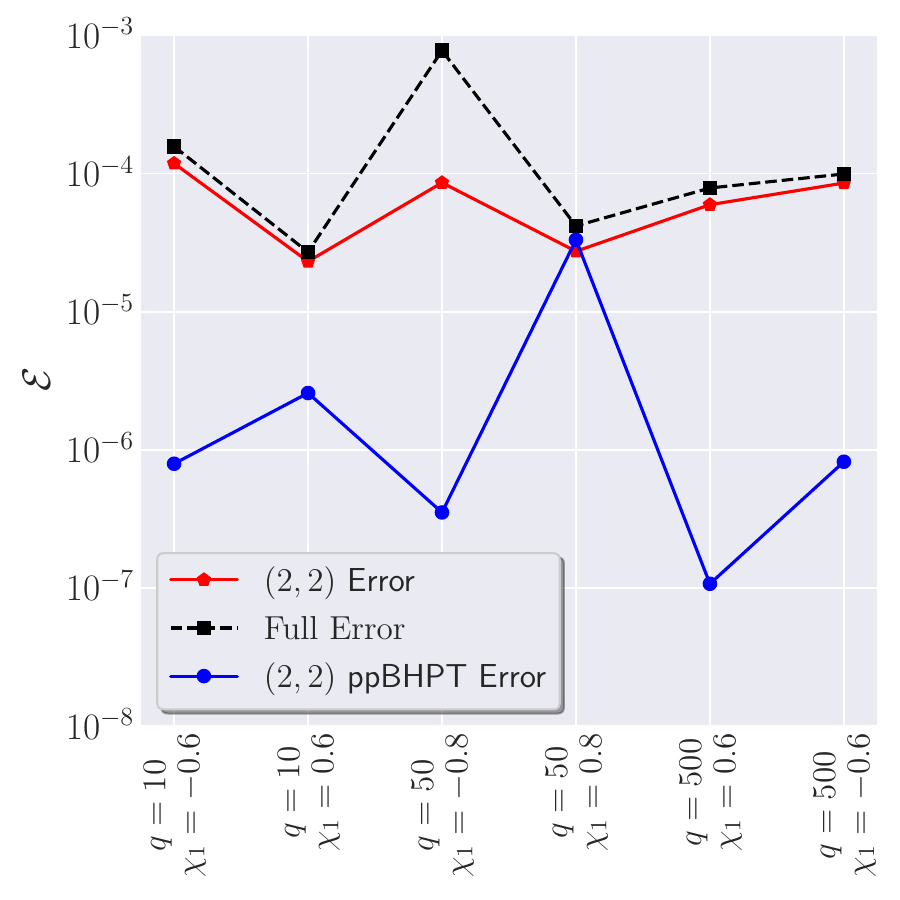}
    \caption{The numerical truncation error (blue circles) estimates the quality of the training data by comparing two numerical solutions of increasing resolution for a particular sampling of $(q, \chi_1)$ across our parameter space. We compare this error to the validation error for the (2,2) mode (red circles) and all modes (black squares).  The data is not continuous, but we have added a connecting line for visual assistance. We find that the surrogate model is able to match the quality of the training data at some points in the parameter space (e.g. $q=50$ and $\chi_1=0.8$) while at other points (e.g. $q=500$ and $\chi_1=0.6$) the error can be orders of magnitude larger. Nevertheless, despite not being able to reproduce the training data's accuracy, the errors archived by our model throughout the parameter space constitute high accuracy (see, for example, Fig.~\ref{fig:loocv_vs_training}).}
    \label{fig:numeric_error}
\end{figure}

To assess the quality of our model three different types of error are calculated. First, we build our surrogate using all 476 ppBHPT waveforms and evaluate the error by comparing the surrogate prediction and the original waveform. This type of error evaluation is called {\em training error}. The training error provides a reasonable measure of the model performance over the parameter space and is straightforward to compute, but it only captures in-sample error. To estimate the model's generalization (out-of-sample) error, we perform leave-one-out cross-validation error, or simply {\em validation error}, where we hold out one training data corresponding to a mass ratio and spin parameter and build a surrogate with the remaining 475 training data points. We evaluate the surrogate at that held-out point in parameter space and compare the predicted data with the corresponding held-out ppBHPT waveform. We repeat this step for all 476 data points, building a conservative error profile across the parameter space. 

These results are displayed in Figures \ref{fig:loocv_full}, \ref{fig:loocv_modes}, and \ref{fig:loocv_vs_training}. Figure \ref{fig:loocv_full} shows the validation error computed using all 8 modeled modes, and Fig. \ref{fig:loocv_modes} shows the mode-by-mode validation error for the quadrupolar mode as well as the most relevant subdominant modes. We calculate the validation error for individual modes using Eq.~\ref{eq:L2} with $(\ell, m)$ fixed to the specific mode. The 95th percentile of the validation error is as follows: for $(\ell,m)= (2,2)$ is $8 \times 10^{-4}$, $(\ell,m)= (2,1)$ is $2.3 \times 10^{-3}$, $(\ell,m)= (3,3)$ is $3.6 \times 10^{-3}$, and $(\ell,m)= (4,4)$ is $8.7 \times 10^{-3}$. In Figure \ref{fig:loocv_vs_training}, we plot frequency density as a function of ${\rm Log}_{10}(\mathcal{E})$ for training error and validation error for the individual modes. Both histograms broadly overlap, indicating that the surrogate model can predict the waveform for an unknown data point without losing quality or overfitting.

Finally, we estimate the numerical resolution error by comparing a handful of representative training waveforms with a high-resolution ppBHPT waveform computed using a weighted essentially non-oscillatory (WENO) solver~\cite{field2023gpu}. In Fig.~\ref{fig:numeric_error}, we compare the validation and numerical solver errors for several data points. For the validation, we show both the $(2,2)$ mode and full model errors. We find that the validation error is often larger than the numerical error, indicating additional training data and/or methodological improvements that could potentially improve our model. Indeed, the aligned-spin model \texttt{NRHybSur3dq8}~\cite{Varma:2018mmi} covering $q \leq 8$ systems is able to achieve accuracies similar to the training data's quality. However, there are differences in the training data (e.g., the appearance of dominant retrograde modes, the larger mass ratios covered, and non-smooth features in the ppBHPT's orbital trajectory) that suggest our modeling problem could be more challenging. Appendix~\ref{app:peaktime} considers numerically finding the waveform's peak as one such challenge associated with our training data.

\begin{figure*}
    \includegraphics[width=\textwidth]{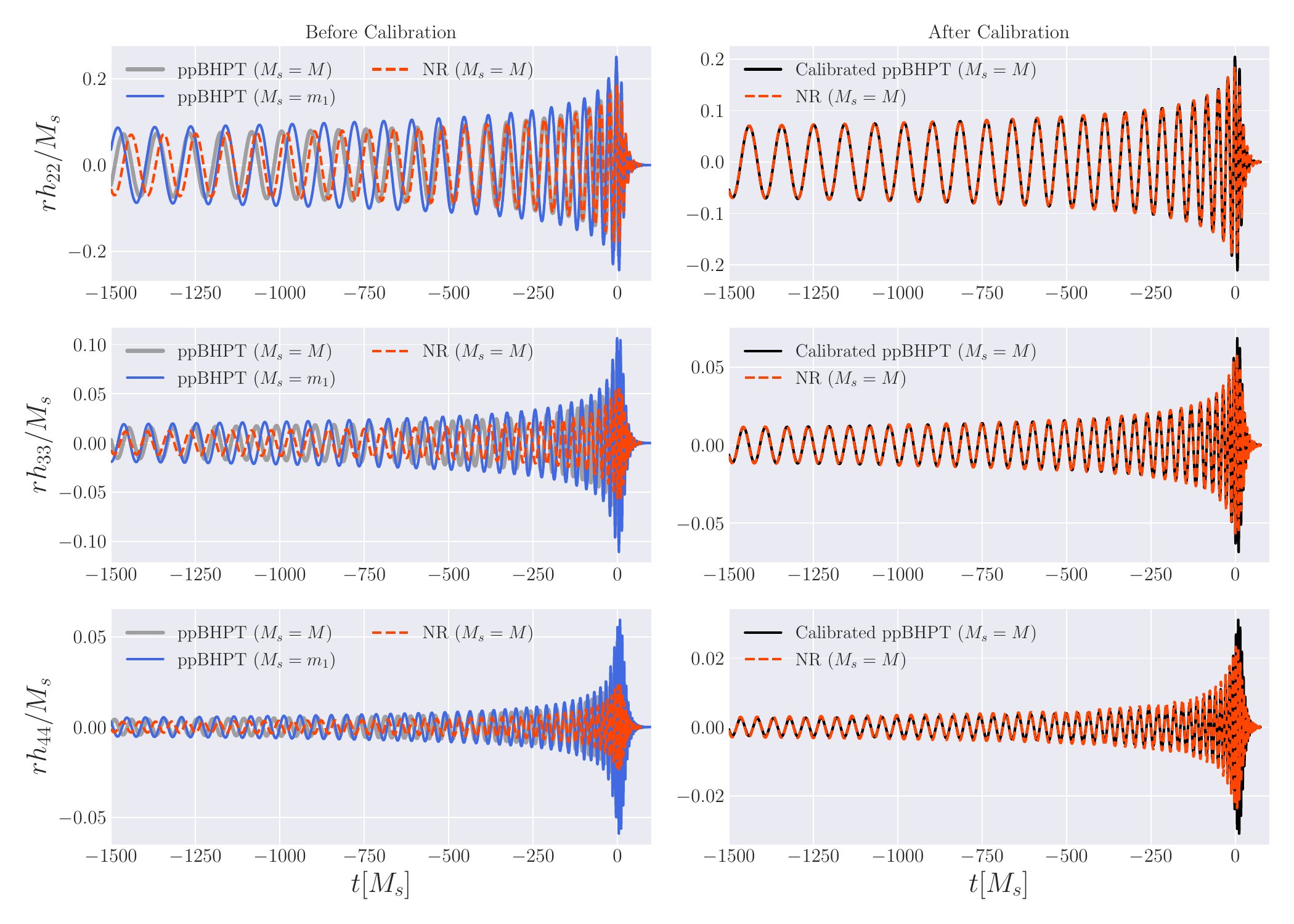}
    \caption{A comparison between waveforms computed by point particle black hole perturbation theory (ppBHPT) to numerical relativity (NR) for a $(q,\chi_1)=(4,0.6)$ system. Shown are the (2,2) mode (top row), the (3,3) mode (middle row), and the (4,4) mode (bottom row). Figures in the left column show the uncalibrated ppBHPT waveforms expressed in terms of a mass scale $M_s=m_1$ commonly used in black hole perturbation theory (blue solid lines) as well as the mass scale $M_s=M$ used in NR (solid grey lines). 
    Figures in the right column show the calibrated ppBHPT waveforms (black solid lines). In both columns we also show the respective NR surrogate model \texttt{NRHybSur3dq8} evaluated at $(q,\chi_1)=(6,0.6)$ (dashed red lines).
    } 
    \label{fig:calibrated}
\end{figure*}

\section{Calibrating perturbation waveforms to numerical relativity}
\label{sec:calibration}

In the previous section, we trained a surrogate model for gravitational waveforms computed within the ppBHPT framework. However, ppBHPT waveforms (and hence our model) faithfully approximate the physically correct ones only in the large mass ratio limit. To construct an accurate model at comparable-to-intermediate mass ratios, we introduce model calibration parameters and set their values with numerical relativity data.

\subsection{Masses and spins in NR and ppBHPT}
\label{sec:parameteriations}

\begin{figure*}[htb]
	\subfigure[]{\label{fig:alpha_22}
		\includegraphics[scale=0.56]{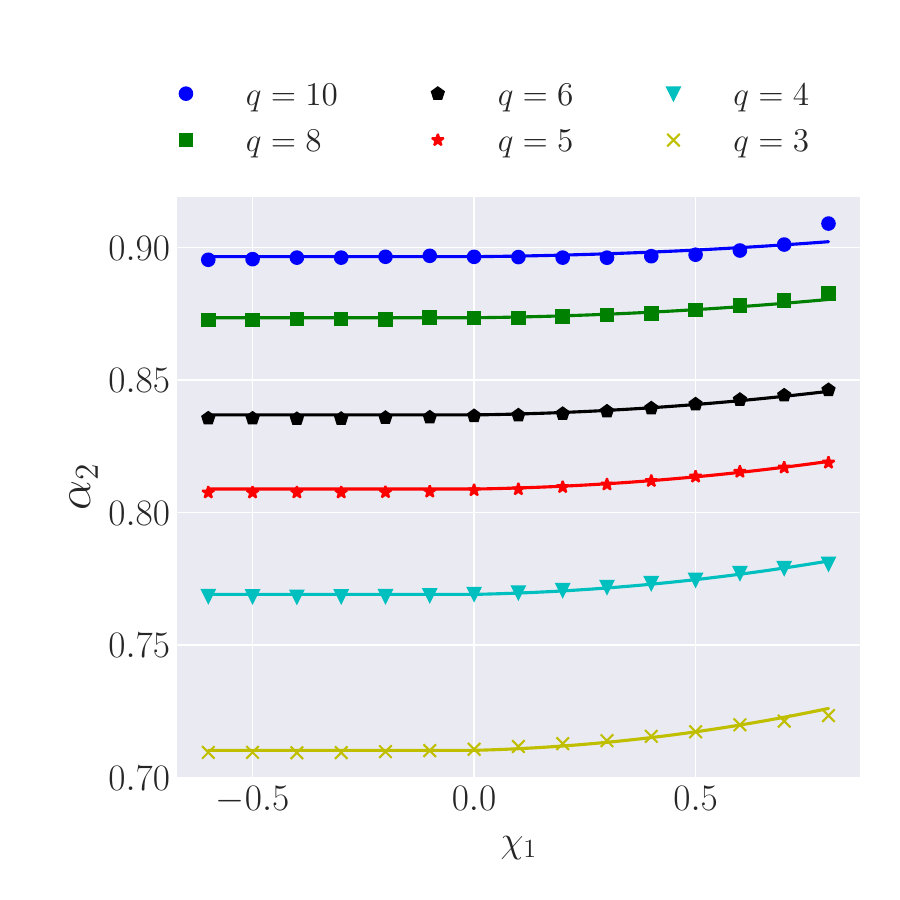}}
	\subfigure[]{\label{fig:beta}
		\includegraphics[scale=0.56]{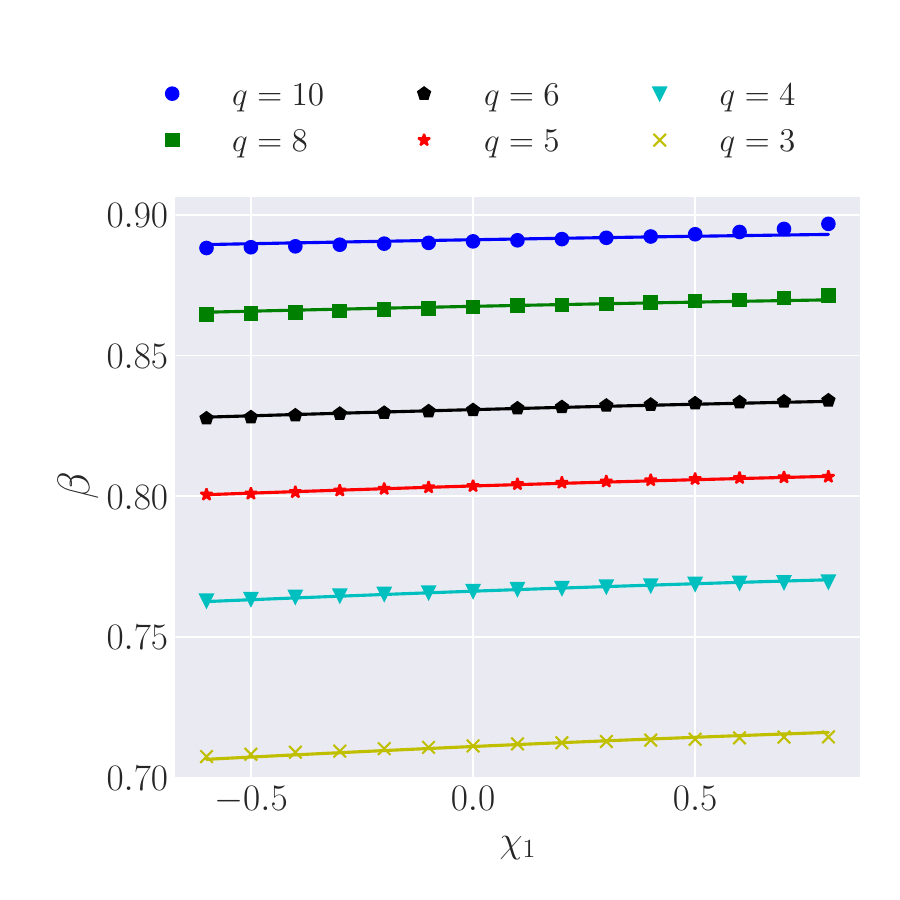}}
	\caption{\label{fig:alpha_beta} (\textit{Left}) The scaling parameters $\alpha_{2}$ obtained by calibrating ppBHPT waveforms to NR for mass ratios in the range $q \in [3,10]$ as a function of the primary black hole's spin $\chi_1$. (\textit{Right}) The scaling parameters $\beta$ obtained by calibrating ppBHPT waveforms to NR for mass ratios in the range $q\in[3,10]$ as a function of the primary black hole's spin $\chi_1$. The solid lines show the regression fits provided from Eqs.~\eqref{alpha_fit} and~\eqref{beta_fit}.
    }
\end{figure*}

Before calibrating with and comparing to NR, we briefly review the different mass and spin parameters used in both settings to facilitate meaningful comparisons. 

In perturbation theory, the background geometry is described by the usual Kerr mass and spin parameters denoted in this subsection as $m_1^\mathrm{PT}$ and  $a_1^\mathrm{PT}$, respectively. Here, we've temporarily added a ``PT'' superscript for clarity; for example, $m_1^\mathrm{PT}$ has precisely the same meaning as $m_1$ already used throughout this paper. The secondary black hole is non-spinning, and its mass parameter is the usual Schwarzschild mass parameter $m_2^\mathrm{PT}$. The waveform computed in perturbation theory is labeled by $q^\mathrm{PT} = m_1^\mathrm{PT} / m_2^\mathrm{PT}$ and $\chi_1^\mathrm{PT} = a_1^\mathrm{PT}/m_1^\mathrm{PT}$, and the mass scale is taken to be $M^\mathrm{PT}_S=m_1^\mathrm{PT}$.

In numerical relativity, following the conventions of the SXS collaboration's catalog~\cite{boyle2019sxs}, waveforms are labeled by quasi-local measurements of each black hole's mass and spin. The masses are the Christodoulou masses of each black hole, $m_1^\mathrm{NR}$ and  $m_2^\mathrm{NR}$, computed on the apparent horizons. The spin of the primary black hole, $\chi_1^\mathrm{NR}$, is also computed on the apparent horizon following the procedure described in Refs.~\cite{Owen:2017yaj,boyle2019sxs}. The waveform computed in NR is labeled by $q^\mathrm{NR} = m_1^\mathrm{NR} / m_2^\mathrm{NR}$, $\chi_1^\mathrm{NR}$, and the mass scale is taken to be $M^\mathrm{NR}_S=m_1^\mathrm{NR} + m_2^\mathrm{NR}$.

The NR-based mass and spin values correspond to the Kerr parameters for a single, isolated black hole. While black holes in an NR simulation are dynamic and not isolated, it is expected that the NR computations of each black hole's mass and spin can be approximated by the Kerr values when the black holes are sufficiently separated, as they happen to be near the start of the simulation when such measurements are made. And so  $q^\mathrm{PT} \approx q^\mathrm{NR}$ and $\chi_1^\mathrm{PT} \approx \chi_1^\mathrm{NR}$.

Clearly, the mass scales of the ppBHPT and NR waveforms are different. This indicates that, even if no other differences between NR and ppBHPT waveforms were to be expected, both the amplitude and time of the ppBHPT waveforms must be scaled by $1/(1+1/q)$, which is the factor needed to convert between a mass-scale of $m_1$ to $m_1 + m_2$. This observation initially motivated the form of the $\alpha$-$\beta$ scaling described in Sec.~\ref{sec:calibration_overview}. However, the approximations of the ppBHPT framework also lead to parameter-dependent modeling errors. We use the calibration parameters to account for these inaccuracies in addition to automatically switching mass scales.

\begin{figure*}[htb]
	\centering
	\subfigure[]{\label{fig:alpha_beta_err}
		\includegraphics[scale=0.5]{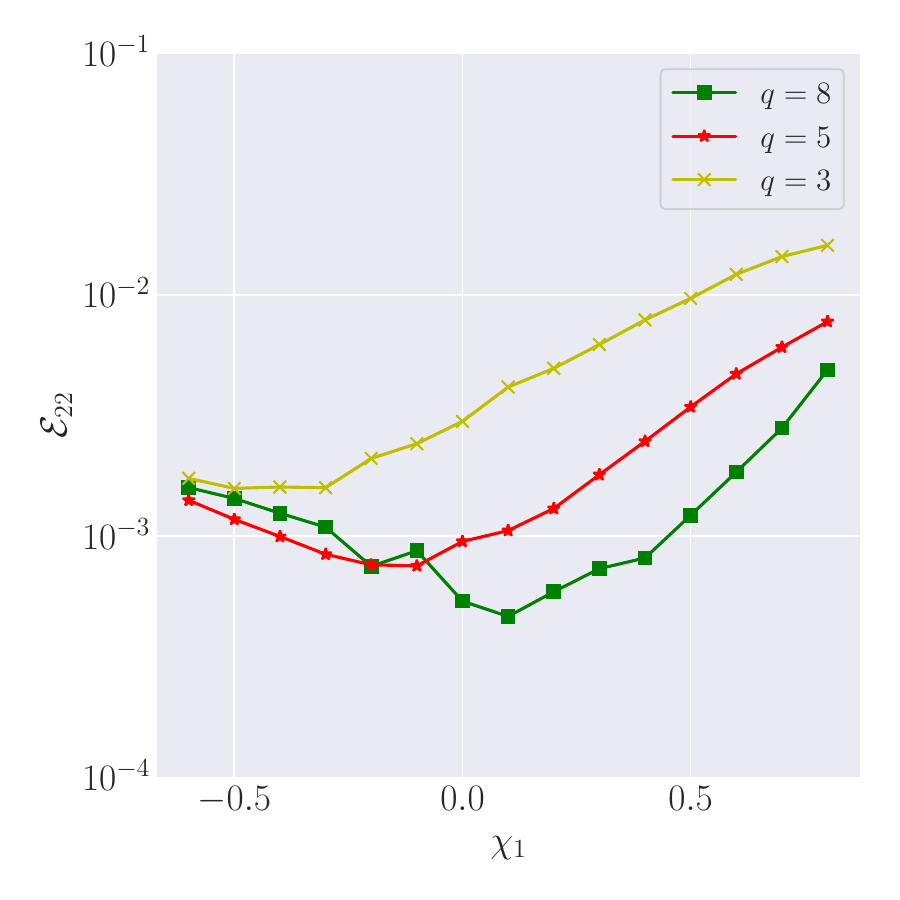}}
	\subfigure[]{\label{fig:2d}
		\includegraphics[scale=0.5]{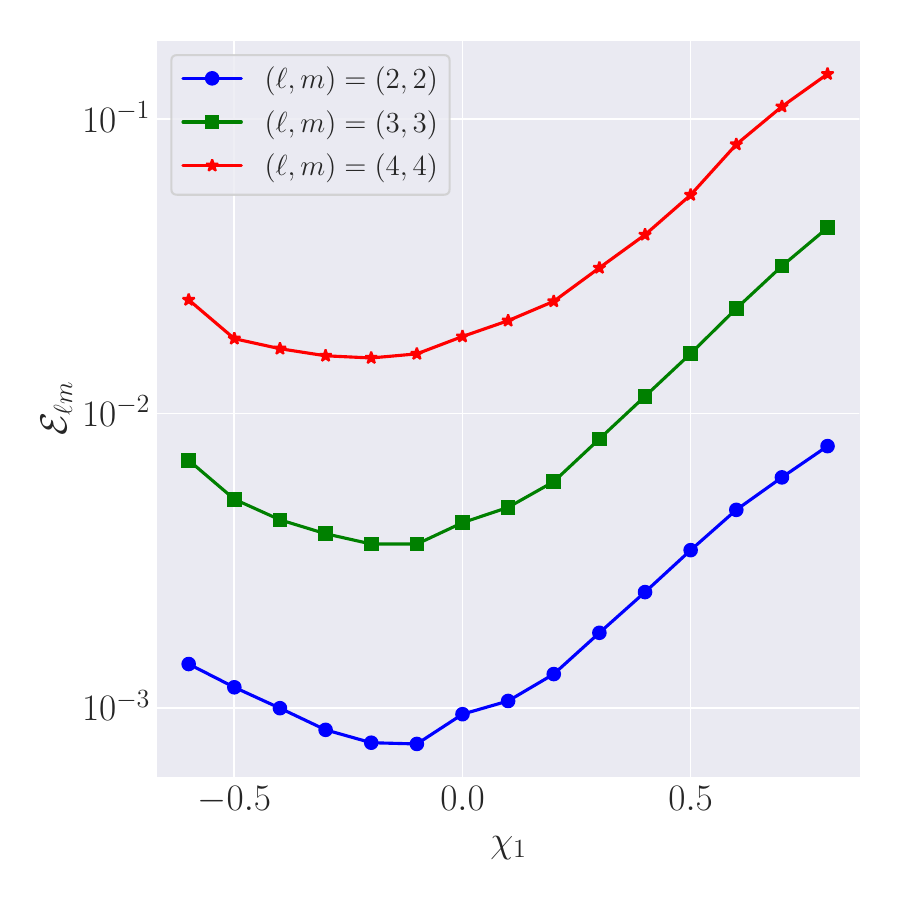}} 
	\caption{\label{fig:err_22_33_44} 
 The time domain $L_2$ norm error between the calibrated ppBHPT waveform's modes and numerical relativity for selected mass ratios in the range $q\in[3,8]$ as a function of primary black hole's spin $\chi_1$. 
(\textit{Left}) Error in modeling the (2,2) mode where the error is measured over the full temporal duration.  (\textit{Right}) Error between the calibrated ppBHPT waveform and the NR surrogate model \texttt{NRHybSur3dq8} for three representative spherical harmonic modes -- the $(2,2)$, $(3,3)$ and $(4,4)$ modes -- as a function of the spin of the primary black hole while fixing $q=5$.
    }
\end{figure*}

\subsection{Overview of the model calibration procedure}
\label{sec:calibration_overview}

Our calibration procedure, which we shall refer to as  $\alpha$-$\beta$ scaling, is identical to the one employed in the non-spinning model \texttt{BHPTNRSur1dq1e4}~\cite{1d_updated}. 
Ref.~\cite{1d_updated} proposed modifying the ppBHPT waveforms,
\begin{equation}
\label{eq:alpha_beta}
h_{\mathrm{S}, \alpha, \beta}^{\ell m}(t ; q, \chi_1)=\alpha_{\ell} h_{\mathrm{S}}^{\ell m}\left(\beta t ; q, \chi_1\right),
\end{equation}
with a single time-independent $\beta(q,\chi_1)$ to scale the time and a set of time-independent $\{\alpha_{\ell}(q,\chi_1)\}$ to scale the amplitudes. 
The values of $\alpha_{\ell}$ and $\beta$ are found by solving an optimization problem,
\begin{equation}
\label{eq:inner_product_alpha_beta}
\min _{\alpha_\ell, \beta} \frac{\int\left|h_{\mathrm{S}, \alpha_\ell, \beta}^{\ell,\ell}(t ; q, \chi_1)-h_\mathrm{NR}^{\ell,\ell}(t ; q, \chi_1)\right|^{2} d t}{\int\left|h_\mathrm{NR}^{\ell,\ell}(t ; q, \chi_1)\right|^{2} d t}\,,
\end{equation}
at discrete parameter values available 
using the \texttt{NRHybSur3dq8} waveform model as a proxy of NR. Note that we use $\ell=m$ modes to obtain $\alpha_{\ell}$ and $\beta$. Values of $\alpha_{\ell}$ are also used to scale the $\ell \neq m$ amplitudes. While there is some $m$-dependence to the amplitude corrections $\alpha_{\ell m}$, the overall model's accuracy is not limited by the $\alpha_{\ell m} \approx \alpha_{\ell \ell}$ approximation. Here $h_{\mathrm{S}}$ is the ppBHPT surrogate waveform, $h_\mathrm{NR}$ is the \texttt{NRHybSur3dq8} waveform model (our proxy~\footnote{As shown in Fig.~6 of Ref.~\cite{varma2019surrogate2}, the \texttt{NRHybSur3dq8} model and NR waveforms are consistent to mismatch errors of $\lesssim 10^{-4}$, thereby justifying its use as a suitable proxy for NR.} of an NR waveform), and the integral is taken over $[-10000M,100M]$. \texttt{NRHybSur3dq8} is an aligned-spin surrogate model built using NR simulations in the mass ratio range $1\le q \le 8$ and for spins $|\chi_{1,2}| \le 0.8$, although the model can be extrapolated up to $q=10$; the distribution of calibration points is shown in Fig.~\ref{fig:alpha_beta_err_heatmap}. Using \texttt{NRHybSur3dq8} in place of NR data does not introduce any systematic error in our study as the numerical errors of the \texttt{NRHybSur3dq8} model are comparable to the numerical truncation errors in NR itself~\cite{Varma:2018mmi}. We reiterate that our scaling parameters adjust for both the mass scale difference (cf. Sec.~\ref{sec:parameteriations}) as well as other missing physics in the ppBHPT setup. In other words, even if NR and ppBHPT produced physically identical waveforms, we would have $\alpha_{\ell}=\beta=1/(1+1/q)$ in such a scenario.

The procedure above provides sample data at discrete $(q, \chi)$ parameter points, and regression techniques (described later on in Sec.~\ref{sec:model_calibration}) are then used to model the calibration parameters' behavior over the parameter space.

\begin{figure}[h!]
    \centering
    \includegraphics[width=8cm]{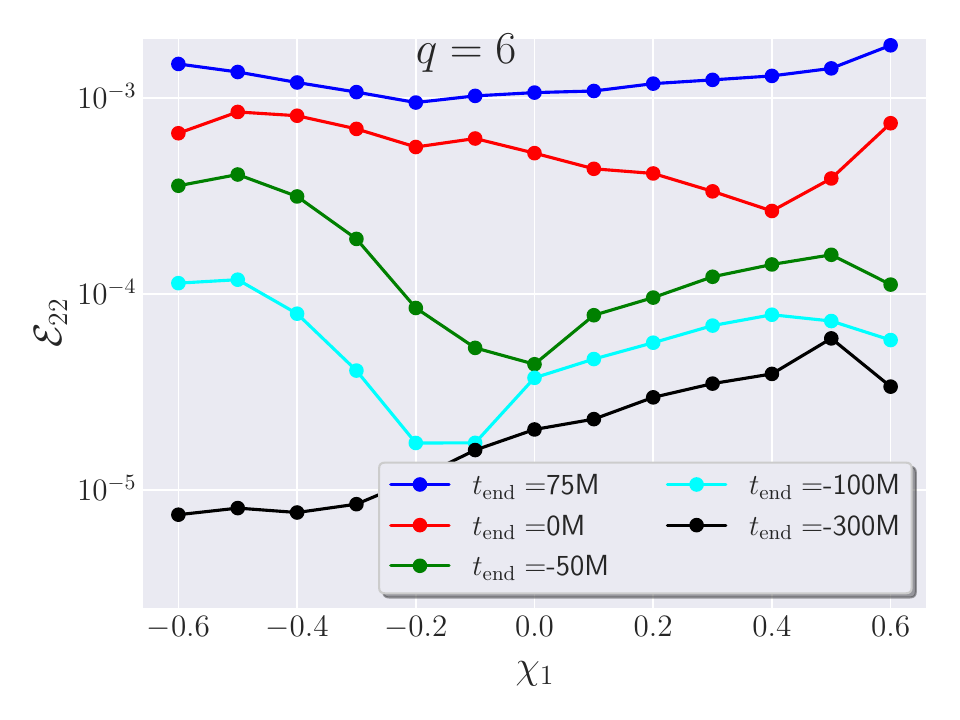}
    \caption{Model calibration results when solving the optimization problem~\eqref{eq:inner_product_alpha_beta}
    over a time window of $[-4500, t_{\rm end}]$. To explore the model's accuracy as a function of $t_{\rm end}$, we show the error between the calibrated ppBHPT waveform's (2,2) mode and numerical relativity for a $q=6$ system. The final time is varied from $t_{\rm end}=-300 \ M$ (inspiral only) to $t_{\rm end}=75 \ M$ (up through ringdown). As discussed in Sec.~\ref{Sec:training_data}, at mass ratios $q \lesssim 20$, the GOT procedure used to connect the inspiral and plunge trajectories can cause unphysical trajectory and waveform features starting at around $\approx -200 M$. These unphysical features are absent in the NR data. A second challenge occurs when attempting to match the near-merger and ringdown signals due to incorrect remnant values, finite size effects, and potentially other complications not captured in traditional ppBHPT.}
    \label{fig:calibration_error_vs_tend}
\end{figure}

\subsection{Model calibration}
\label{sec:model_calibration}

\begin{figure}[h!]
    \centering
    \includegraphics[width=8cm]{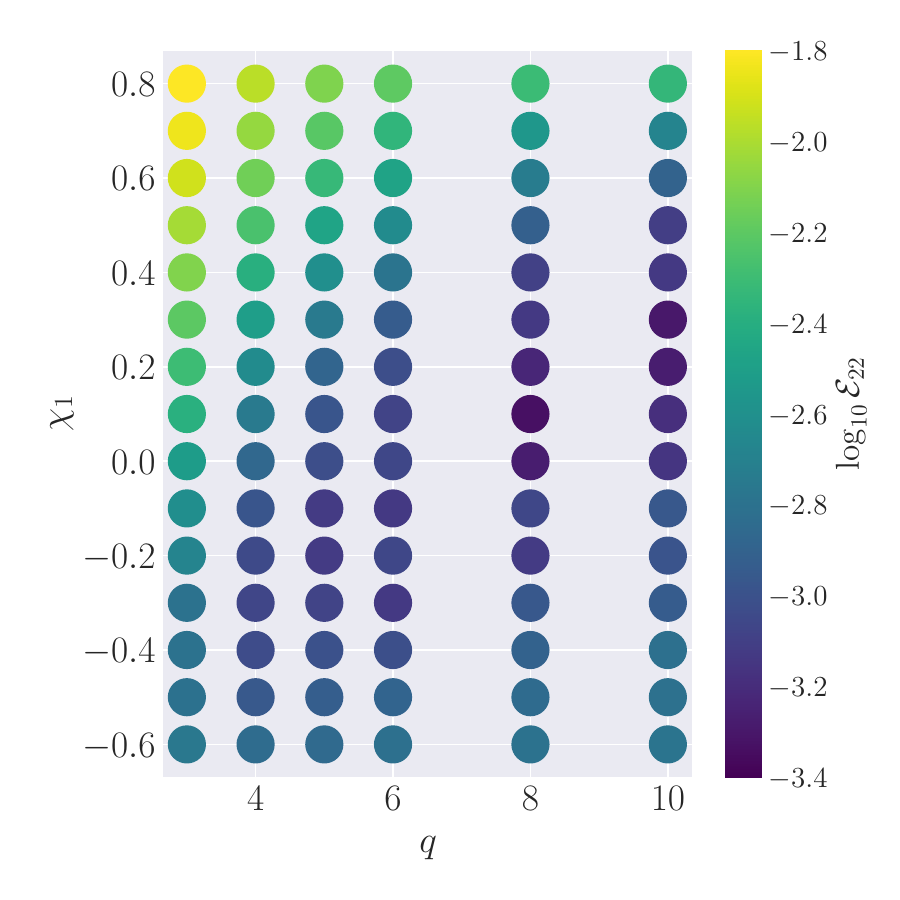}
    \caption{Heatmap of the time domain relative $L_2$ norm error between the calibrated ppBHPT waveforms and the NR surrogate model \texttt{NRHybSur3dq8} (as a proxy of NR) as a function of the mass ratios $q$ and the spin of the primary black hole $\chi_1$ in the comparable mass regime. 
    }
    \label{fig:alpha_beta_err_heatmap}
\end{figure}

To model each calibration parameter's $q$ and $\chi_1$ dependence, we sample from $q=3$ to $q=10$ and vary the spin from $\chi_1=-0.6$ to $\chi_1=0.8$, giving a total of $90$ data points as shown in Fig.~\ref{fig:alpha_beta_err_heatmap}.  These data are then used to fit $\alpha_{\ell}$ and $\beta$ to polynomials in $1/q$ and $\chi_1$, 
\begin{align}
\label{alpha_fit}
    \alpha_{\ell}(q,\chi_1) & = 1 +  
    (\sum_{n=1}^4 A_{\alpha_\ell}^{n} q^{-n}) 
    (1+\sum_{n=1}^2 B_{\alpha_\ell}^n \chi_1^{n}) \,, \\
    \beta(q, \chi_1) & = 1+ (\sum_{n=1}^4 A_{\beta}^n  q^{-n})
    (1+\sum_{n=1}^2 B_{\beta}^n\chi_1^n) \,,
\label{beta_fit}
\end{align}
where the polynomial order has been chosen to yield the smallest fitting errors. The values of these coefficients are provided in Tables \ref{Tab:alpha_values} and \ref{Tab:beta_values}. Note that, as we see little changes in $\alpha_3$ and $\alpha_4$ as we change spin, we fix $B_{\alpha_3}^n$ and $B_{\alpha_4}^n$ to be zero. We show the fit predictions as solid lines in Fig.~\ref{fig:alpha_22} and Fig.~\ref{fig:beta}.

In Figure \ref{fig:calibrated}, we demonstrate the effectiveness of the calibration procedure for a system with mass ratio $q=6$ and spin $\chi_1=0.6$. 
We show both the ppBHPT (blue solid lines) and NR (red dashed lines) waveforms before and after scaling the $(2,2)$, $(3,3)$, and $(4,4)$ modes. Before the calibration, neither the amplitudes nor the phasing match, giving a relative error -- computed using Eq.(\ref{eq:L2}) -- of about $1$. We first show the ppBHPT waveforms scaled by only $1/(1+1/q)$, a rough guess for $\alpha_{\ell}$ and $\beta$ based on mass scale considerations only. We find that ppBHPT waveforms scaled by $1/(1+1/q)$, while visually closer to NR, are still noticeably different. However, upon scaling the waveforms with values of $\alpha_{\ell}$ and $\beta$ found through solving the optimization problem~\eqref{eq:inner_product_alpha_beta}, both waveforms visually match mode-by-mode. The relative $L_2$-norm error~\footnote{Applying Eq.~\eqref{eq:L2} where the sum in both the numerator and denominator is taken over a single mode.} between the calibrated ppBHPT and NR waveforms is 0.002 for the $(2,2)$ mode, 0.001 for the $(3,3)$ mode, and 0.005 for the $(4,4)$ mode. 

To verify the effectiveness of the $\alpha$-$\beta$ scaling everywhere in the parameter space and to understand the accuracy of the scaled waveforms, we repeat the scaling procedure for 90 points in the parameter space. These points are chosen as a subset of the training data (cf. Fig.~\ref{fig:training_bank}) such that the mass ratio and spin ranges satisfy $3 \le q \le 10$ and $-0.6 \le \chi_1 \le 0.8$. This range is selected as its where we have an accurate NR surrogate model \texttt{NRHybSur3dq8} ($q \le 10$) and to avoid complicated signal morphology as discussed in Sec.~\ref{app:retrograde} ($\chi_1<-0.6$). The distribution of our NR calibration data, as well as the calibration error at these points, is shown in Fig.~\ref{fig:alpha_beta_err_heatmap}. In particular, this empirically demonstrates that a time-independent $\beta(q,\chi_1)$ and $\{\alpha_{\ell}(q,\chi_1)\}$ can be used to scale the ppBHPT waveform's amplitude and phase to make it approximately match an NR waveform.

Figures \ref{fig:alpha_22} and \ref{fig:beta} show the calibration parameters $\alpha_2$ and $\beta$ behavior throughout the parameter space. We find that for a fixed spin, as the mass ratio increases, both $\alpha_2$ and $\beta$ approach one. This is expected as the ppBHPT framework becomes exact as $q\rightarrow \infty$; hence, the calibration parameters must approach unity in this limit. Interestingly, the calibration parameters show only a mild dependence on the spin. Furthermore, $\alpha_2$ and $\beta$ remain almost constant for negative spin values. The other calibration parameters ($\alpha_3$ and $\alpha_4$; not shown) show a weaker dependence on spin. 

Finally, in Figure \ref{fig:alpha_beta_err}, we show the $L_2$-norm error between the between the scaled ppBHPT and NR waveform's dominant quadrupolar mode. The errors are typically smaller than $10^{-2}$, indicating a reasonable accuracy across the parameter space $q \leq 8$. At higher mass ratios, the ppBHPT framework becomes a more faithful approximation of the problem, and we expect better agreement. While there is a lack of high mass ratio NR waveforms to perform an exhaustive check, in Sec.~\ref{sec:validation_high_q} we show good agreement to NR at $q=15$.

We then repeat this analysis with the higher modes and find that the higher modes exhibit larger errors, which was also observed in the previous \texttt{BHPTNRSur1dq1e4} model. In Fig.~\ref{fig:2d}, we show the errors for three different spherical harmonics modes for different spin values while keeping the mass ratio value fixed at $q=5$. For example, $(3,3)$ mode errors are almost one order of magnitude larger than the $(2,2)$ mode. However, the errors are still of the order of $\sim 10^{-2}$. The largest errors we encounter in our calibration are for the $(4,4)$ mode, where errors can reach as large as $\sim 10^{-1}$ for $\chi_1=0.8$. 

To check whether the errors behave differently in the inspiral and merger-ringdown part, in Fig.~\ref{fig:calibration_error_vs_tend}, we show the relative differences between \model and \texttt{NRHybSur3dq8} when the upper range of the calibration window is varied over the late inspiral, plunge, merger, and ringdown regimes. We note that the inspiral-only error is two orders of magnitude smaller than the full waveform error, which suggests the latter portions of the ppBHPT waveform cannot be as accurately calibrated to NR using our procedure. There are potentially multiple reasons for this behavior. First, as discussed in Sec.~\ref{Sec:training_data}, at mass ratios $q \lesssim 20$, the GOT procedure used to connect the inspiral and plunge trajectories can cause unphysical trajectory and waveform features starting at around $\approx -200 M$. These unphysical features are absent in the NR data. A second challenge occurs when attempting to match the near-merger and ringdown signals due to incorrect remnant values, finite size effects, and potentially other complications not captured in traditional ppBHPT.

\begin{figure}
    \centering
    \includegraphics[width=0.45\textwidth]{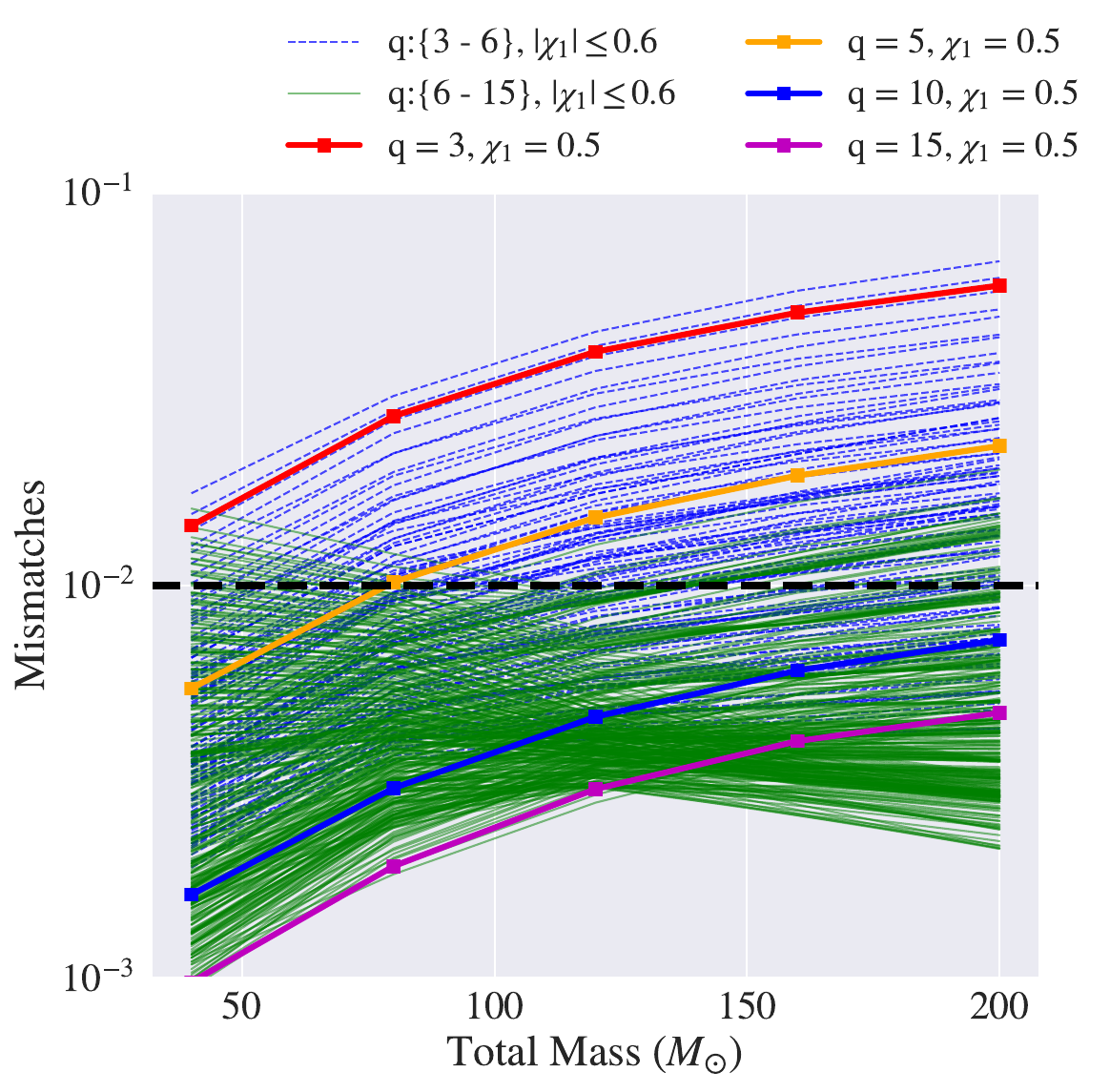}
    \caption{Frequency-domain mismatches between the NR-calibrated ppBHPT surrogate model \model and \texttt{NRHybSur2dq15} (as a proxy of NR) for mass ratios and spins covering a range of $3 \leq q \leq 15$ and $-0.6 \leq \chi_1 \leq 0.6$, respectively. The mismatches are shown as a function of the binary total mass $M$ at inclination $\iota=0.0$ and orbital phase $\phi=0.0$, and are computed using the advanced LIGO design sensitivity noise curve. The dashed horizontal line demarcates a mismatch of 0.01, a commonly used threshold for sufficiently good model quality. As the \model model is more accurate in the pre-merger regime, we generally find: (i) mismatches decrease for lighter systems (which have more in-band inspiral cycles) and (ii) mismatches are generally better for the higher range of mass ratios (solid green lines) compared to the lower range (blue dashed lines) calculated here. The solid lines with square markers} show the mismatches for a sequence of $\chi_1=0.5$ systems where the mass ratio is systematically varied from $3$ to $15$.
    \label{fig:mismatches_spin}
\end{figure}


\subsection{Frequency domain error between NR and calibrated ppBHPT}

\begin{figure*}
    \includegraphics[width=\textwidth]{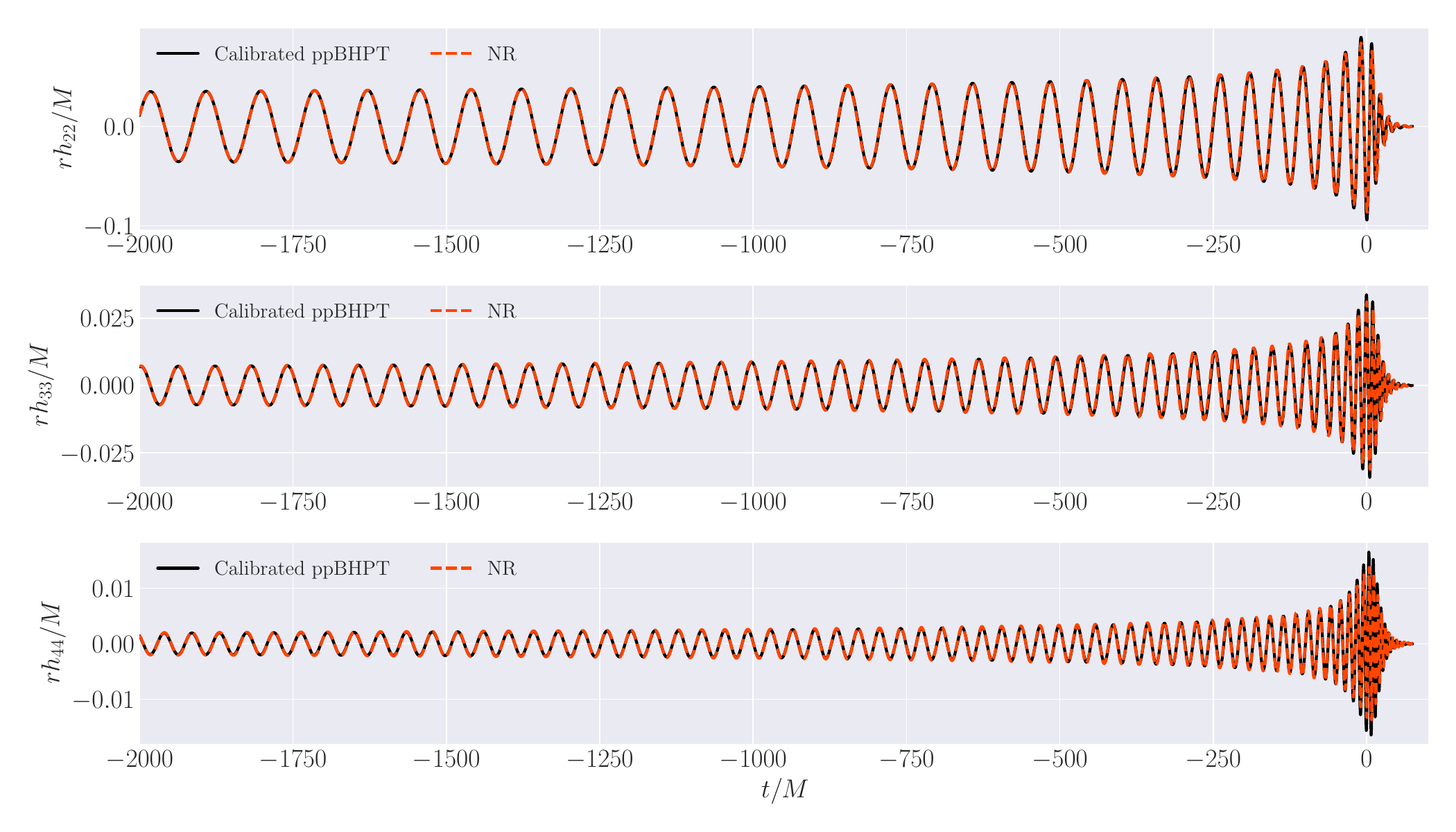}
    \caption{NR surrogate model \texttt{NRHybSur2dq15} waveform (red dashed line; labeled as ``NR'') and the representative scaled ppBHPT surrogate waveform (black solid lines) at $(q,\chi_1)=(15,0.4)$ for three representative modes $(2,2)$, $(3,3)$, and $(4,4)$.
    }
    \label{fig:q15_0.4}
\end{figure*}

\begin{figure*}
    \includegraphics[width=\textwidth]{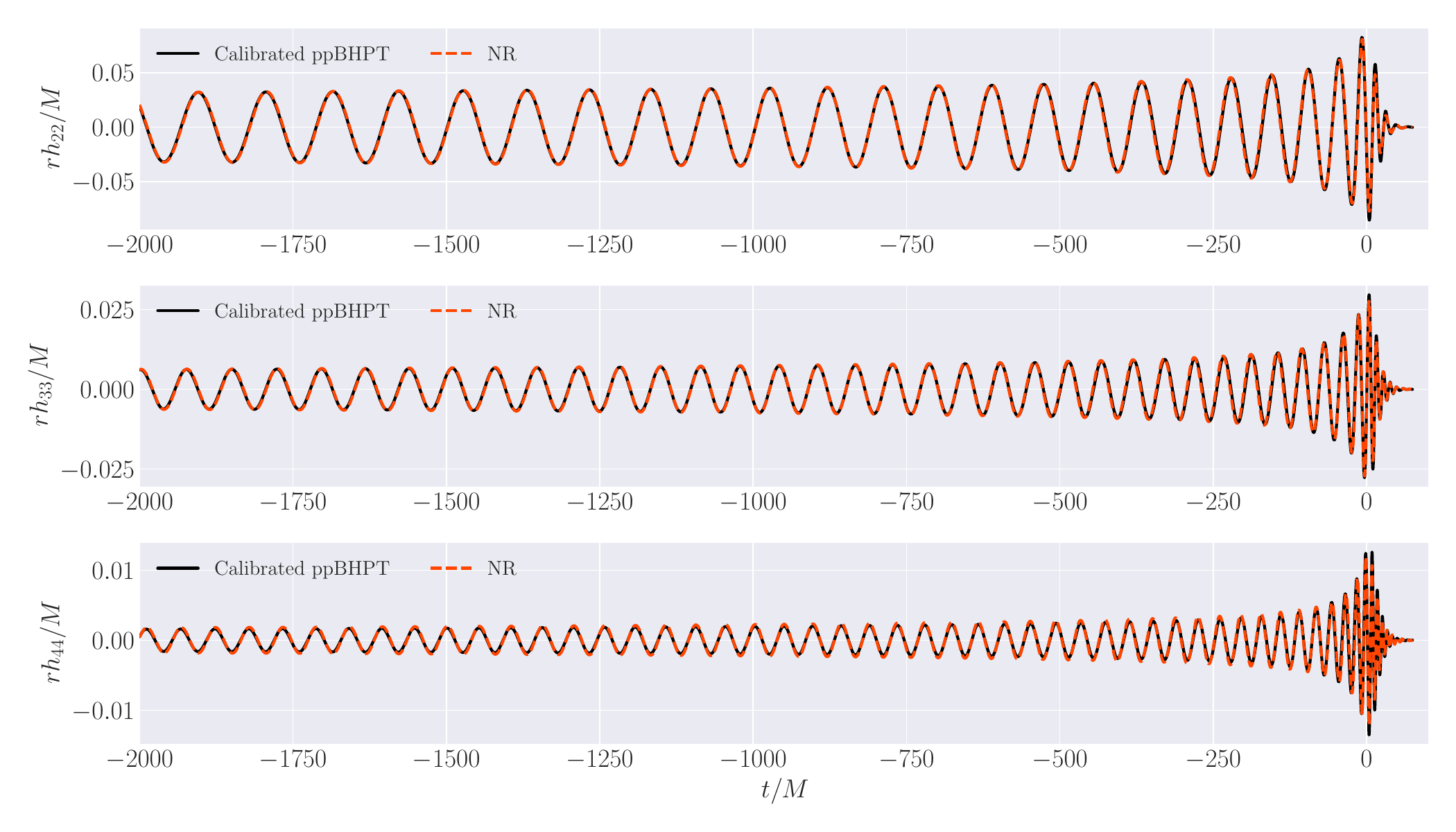}
    \caption{NR surrogate model \texttt{NRHybSur2dq15} waveform (red dashed line; labeled as ``NR'') and the representative scaled ppBHPT surrogate waveform (black solid lines) at $(q,\chi_1)=(15,-0.4)$ for three representative modes $(2,2)$, $(3,3)$, and $(4,4)$.}
    \label{fig:q15_n0.4}
\end{figure*}

To assess the model's accuracy for data analysis purposes, we compute the frequency domain mismatches between \model and the NR surrogate model \texttt{NRHybSur2dq15}~\cite{yoo2022targeted}. We use \texttt{NRHybSur2dq15} in particular since the model's accuracy is comparable to NR for $q \leq 15$ and $|\chi_1| \leq 0.5$. The frequency domain mismatch between two waveforms $h_1$ and $h_2$ is defined as,
\begin{gather}
	\left<h_1, h_2\right> = 4 \mathrm{Re}
	\int_{f_{\mathrm{min}}}^{f_{\mathrm{max}}}
	\frac{\tilde{h}_1 (f) \tilde{h}_2^* (f) }{S_n (f)} df,
	\label{Eq:freq_domain_Mismatch}
\end{gather}
where $\tilde{h}(f)$ indicates the Fourier transform of the strain $h(t)$, $^*$ indicates complex conjugation, $\mathrm{Re}$ indicates the real part, and $S_n(f)$ is the one-sided power spectral density of the Advanced LIGO detector at its design sensitivity. Before Fourier transforming the time-domain waveform to the frequency domain, we first (i) taper the time domain waveform using a Planck window~\cite{McKechan:2010kp}, (ii) zero-pad to the nearest power of two, and finally (iii) increase the length of the waveform by a factor of 8 by zero-padding. Tapering at the start of the waveform is done over $1.5$ cycles of the $(2,2)$ mode while tapering at the end is done over the last $20M$. We set $f_{\rm min}$ to be twice the orbital angular velocity at the end of the first tapering window while $f_{\rm max}$ is chosen to be eight times the angular velocity measured at the peak of the waveform. This choice allows us to resolve up to the $m=4$ mode. Following the procedure described \cite{Blackman:2017pcm}, the mismatches are optimized over shifts in time, polarization angle, and initial orbital phase. The plus and cross polarizations are handled on equal ground by employing a two-detector setup, where one detector observes exclusively the plus polarization and the other observes the cross-polarization.

In Fig. \ref{fig:mismatches_spin}, we show the frequency domain mismatches for a total of 400 points in the intrinsic parameter space, uniformly spanning mass ratios from $q=3$ to $q=15$ and spins from $\chi_1=-0.6$ to $\chi_1=0.6$. Blue dash lines represent mismatches for systems with $3 \leq q \leq 6$, while the green solid lines depict mismatches for systems with $6 \leq q \leq 15$ cases. Generally, near-comparable mass ratios systems (blue dashed lines) exhibit higher mismatches as compared to high mass ratio systems (solid green lines). This is expected as our perturbation theory-based model generally performs better at higher mass ratios. We also observe that lighter systems show smaller mismatches than heavier systems. This is also expected as the late-inspiral through ringdown portions of the signal are generally harder to model. Additionally, we observe that as the mass ratio increases, the mismatches typically decrease, as shown by the sequence of mismatches for systems with $\chi_1 = 0.5$ (solid lines with markers). The horizontal black dashed line indicates a typical mismatch threshold of 0.01, often used in detection and data analysis. We see that the \model model satisfies this criterion for nearly all systems with $q\gtrsim 6$, underscoring that our model is most applicable for large mass ratio systems, precisely where other models are less well developed.

\subsection{Validation against NR at high mass ratio}
\label{sec:validation_high_q}

We now validate our model predictions against an NR surrogate model \texttt{NRHybSur2dq15} that has been trained on high mass ratio NR simulations at different spins values.
We evaluate the NR surrogate model for spins $|\chi_1| \leq 0.4$ and $q=15$ and compare them against the scaled ppBHPT surrogate model. We focus only on the last $5000M$ of the surrogate output as it is trained directly from the NR data over this time interval. Over this range of spins, we find that the $L_2$-norm error for the $(2,2)$ mode is about $\sim 0.001$ while for the subdominant modes, we model around $\sim 0.01$. This demonstrates the effectiveness of \model at high mass ratios. In Fig.~\ref{fig:q15_0.4}, we show the waveform from both the NR surrogate model \texttt{NRHybSur2dq15} and scaled ppBHPT at $(q,\chi_1)=(15,0.4)$ whereas Fig.~\ref{fig:q15_n0.4} shows a similar comparison at $(q,\chi_1)=(15,-0.4)$.

\section{Summary}
\label{sec:summary}

In this paper, we have developed a new surrogate model, \model, covering the comparable-to-intermediate mass ratio regime ($q \leq 10^3$) with aligned spin configurations ($\left| \chi_1 \right| \leq 0.8$, $\chi_2=0$) and mode content spanning $(\ell, m) = \{(2, \{2,1\}), (3,\{3,2,1\}), (4,\{4,3,2\})\}$. As described in Sec.~\ref{Sec:buildSurrogate}, we train this model on waveforms found through numerically solving the Teukolsky equation sourced by a test particle with an adiabatically-driven inspiral. 

This model can faithfully reproduce point-particle black hole perturbation theory (ppBHPT) waveforms with median and 95th percentile time-domain mismatch errors (computed according to Eq.~\eqref{eq:L2}) of $8.3\times 10^{-5}$ and $9.8\times10^{-4}$, respectively. However, the validity of the ppBHPT approximation decreases as it is applied to lower and lower mass ratios. Our model is, therefore, most reliable in the large mass ratio limit.

To construct an accurate model at comparable-to-intermediate mass ratios, we introduce model calibration parameters, $\alpha$ and $\beta$, and set their values by fitting to data from a numerical relativity hybrid surrogate model. Our resulting calibrated surrogate model \model is expected to faithfully reproduce waveforms in the intermediate mass ratio region of the parameter space. While there is a lack of intermediate mass ratio NR waveforms to perform an exhaustive check, in Sec.~\ref{sec:validation_high_q} we show good agreement to NR at $q=15$, indicating our model primarily targets IMRI systems when high accuracy is needed. 

As compared to previous models based on ppBHPT waveforms calibrated to NR, the new model \model allows for (i) spin on the primary black hole, (ii) extends the NR calibration technique to include spin, (iii) and leverages domain decomposition techniques (in time and parameter space) to handle the excitation of retrograde and prograde quasi-normal modes in systems (explored more fully in Appendix~\ref{app:retrograde}) with significant negative spin. These techniques allow for an accurate representation of the waveform's inspiral and merger-ringdown phases. We assessed the model's accuracy by comparing with NR waveforms up to mass ratios of $q=15$ (cf. Figs.~\ref{fig:err_22_33_44}, \ref{fig:mismatches_spin}, \ref{fig:q15_0.4}, \ref{fig:q15_n0.4}), finding the errors generally decrease at higher mass ratios. It is striking that a simple two-parameter ($\alpha$ and $\beta$) calibration captures the physical features necessary to build an accurate model of the waveforms using ppBHPT. Moreover, it is also quite remarkable that $\alpha$ and $\beta$ depend very weakly on spin (cf. Fig.~\ref{fig:alpha_beta}). 

Looking ahead, we aim to extend our model further by incorporating orbital eccentricity, increasing the total duration allowed by the model, and handling inclined orbits. Such extensions will be crucial for capturing the full diversity of intermediate-to-extreme mass ratio binary black hole mergers observable by current and future gravitational wave detectors. \model will be publicly available as part of both the Black Hole Perturbation Toolkit and GWSurrogate.

\noindent{\em Acknowledgments} -- This work makes use of, and contributes to, the Black Hole Perturbation Toolkit. We thank Jonathan Blackman for his early development of the Python package PySurrogate, which is used for portions of the model building. We thank Collin Capano and Leo Stein for insightful discussions on quasi-normal modes. The authors acknowledge the support of NSF grants PHY-2207780 (K.R.), PHY-2307236 (G.K.), PHY-2110496 (S.F.), DMS-2309609 (T.I., S.F., and G.K.), PHY-2110384 (S.A.H.), PHY-2309301 (V.V.) and UMass Dartmouth's Marine and Undersea Technology (MUST) research program funded by the Office of Naval Research (ONR) under grant no. N00014-23-1-2141 (S.F. and V.V.). K.R. is a member of the Weinberg Institute and this manuscript has preprint number UT-WI-23-2024. Most of this work was conducted on the UMass-URI UNITY supercomputer supported by the Massachusetts Green High-Performance Computing Center (MGHPCC) and CARNiE at the Center for Scientific Computing and Data Science Research (CSCDR) of UMassD, which is supported by  ONR/DURIP grant no.\ N00014181255. 


\begin{table*}[htb]
	\centering
    \begin{subtable}
        \centering
        \setlength{\tabcolsep}{10pt}
        \begin{tabular}{l|l|l|l|l|l|l}
            \toprule
            $\ell$ &$A_{\alpha_{\ell}}^{n=1}$ &$A_{\alpha_{\ell}}^{n=2}$ &$A_{\alpha_{\ell}}^{n=3}$ &$A_{\alpha_{\ell}}^{n=4}$ &$B_{\alpha_{\ell}}^{n=1}$ &$B_{\alpha_{\ell}}^{n=2}$\\
            \hline
            2 &-1.15397324 &1.48758115 &-3.35617643 &4.36611547 &-0.01502512 &-0.06650368\\
            3 &-2.70721357 &3.45771825 &-4.26626015 &5.48741687 &0.0 &0.0\\
            4 &-3.22349039 &2.97668803 &5.98484158 &-13.090528 &0.0 &0.0\\
            \botrule	
        \end{tabular}
        \caption{Fitting coefficients for $\alpha_{\ell}$ parameters defined in Eq.(\ref{alpha_fit}).}
        \label{Tab:alpha_values}
    \end{subtable}

    \vspace{1em} 

    \begin{subtable}
        \centering
        \setlength{\tabcolsep}{10pt}
        \begin{tabular}{l|l|l|l|l|l}
            \toprule
            $A_{\beta}^{n=1}$ &$A_{\beta}^{n=2}$ &$A_{\beta}^{n=3}$ &$A_{\beta}^{n=4}$ &$B_{\beta}^{n=1}$ &$B_{\beta}^{n=2}$\\
            \hline
            -1.21099811 &1.31265337 &-0.8174404 &-0.073364777 &-0.02433145 &0.00328884\\
            \botrule	
        \end{tabular}
        \caption{Fitting coefficients for $\beta$ parameters defined in Eq.(\ref{beta_fit}).}
        \label{Tab:beta_values} 
    \end{subtable}
    
\end{table*}


\appendix
\section{Excitation of prograde and retrograde QNMs in the ringdown signal}
\label{app:retrograde}

In Fig.~\ref{fig:ringdown}, the ppBHPT ringdown data shows non-trivial amplitude modulations and phase reversal behavior for binaries with $\chi_1<-0.6$. These features pose significant modeling challenges, for which domain decomposition has been helpful (cf. Sec.~\ref{sec:retrograde_modes}). Recent work has shown that in black hole perturbation theory, the prograde and retrograde modes are both generically excited for retrograde orbits (when the orbital angular momentum points in a direction opposite the black hole's spin)~\cite{lim2019exciting, Hughes:2019zmt}. For numerical relativity simulations, typically restricted to mass ratios $q\leq 4$ or $q\leq 8$, it has also been found that including both prograde and retrograde modes can sometimes lead to better models of the ringdown signal~\cite{dhani2021importance,finch2021modeling,zertuche2022high,zhu2023black,forteza2020spectroscopy}. 

This appendix shows that many of our training waveforms (namely those from BBH systems with large, negative spin) include considerable excitations of both flavors of quasi-normal modes (QNMs) such that the waveform's phase reverses direction. Furthermore, by considering a recent numerical relativity simulation of a BBH system with $(q,\chi_1)=(15,-0.5)$, we consider the extent to which perturbation theory applied at $q=15$ qualitatively matches the numerical relativity ringdown signal. Unlike previous studies~\cite{dhani2021importance,finch2021modeling,zertuche2022high,zhu2023black,forteza2020spectroscopy}, our system's large mass ratio and negative spin suggest that both prograde and retrograde modes should be more readily excited. 

A generic ringdown signal can be expressed as an infinite sum of damped sinusoids known as quasi-normal modes. Following the notation of Ref.~\cite{isi2021analyzing}, each QNM, $\omega_{\ell m n}^p$, is labeled by four indices $(p, \ell, m, n)$, where $(\ell, m)$ are the usual harmonic indices, $n$ is the overtone index, and $p= \mathrm{sgn}(m \mathrm{Re}\left[\omega_{\ell m n}^p\right])$ takes on possible values of $p = \{+,-\}$. For systems where the black hole's spin is aligned with the z-axis (i.e., $\chi_1 \geq 0$), the value of $p$ is typically associated with perturbations co-rotating ($p=+$) or counter-rotating ($p=-$) with the black hole. In our case, we allow the black hole's spin to take on positive and negative values. Since the orbital angular momentum always aligns with the z-axis, when $\chi_1< 0$ the orbital angular momentum and spin vectors are anti-parallel. Most other QNM studies, by comparison, assume $\chi_1>0$ (or, when working with numerical relativity data, $\chi_f >0$), which can be enacted as a post-processing step on the waveform data by reflecting about the equatorial plane ($\theta \rightarrow \pi - \theta$). However, this appendix aims to highlight new waveform modeling difficulties due to exciting both $p=+$ and $p=-$ flavors of QNMs, and we do not reflect about the equatorial plane when modeling waveforms from retrograde orbits. Hence, we do not perform any such coordinate transformation here. 

When discussing our $h_{22}(t)$ simulation data, we will assume the ringdown signal's $(2,2)$ spherical harmonic mode can be modeled by
\begin{align}
    h_{22}(t) = A_{2,2,0}^{+} \exp(-\mathrm{i} \omega_{220}^{+})
    +  A_{2,2,0}^{-} \exp(-\mathrm{i} \omega_{220}^{-}) \,,
\end{align}
which neglects mode mixing and overtones. These contributions are expected to be small and can be safely ignored to demonstrate the excitation of QNMs with both positive and negative frequencies~\cite{giesler2019black,finch2021modeling,zertuche2022high}. Given our non-standard setup that allows for $\chi_1<0$, we will avoid using the terms~\footnote{Indeed, when $\chi_1<0$ these terms would be reversed. For example, when $m=2$, the positive-frequency ($p=+$) QNM is counter-rotating with the black hole (ie a retrograde mode).} ``prograde'' and ``retrograde'' in favor of positive-frequency QNMs ($p=+$ when $m>0$) and negative-frequency QNMs ($p=-$ when $m>0$). Finally, we note that the relevant eigenvalue problem defining the complex values $\omega_{\ell m n}^p$ is often posed for $\chi_1 \geq 0$. Values of  $\omega_{\ell m n}^p$ for $\chi_1 \leq 0$ can be readily obtained from the corresponding  $\omega_{\ell m n}^p(\chi_1 \geq 0)$ ones after making the replacement $m \rightarrow -m$.

To our knowledge, with the notable exception of Ref.~\cite{Lim:2022veo}, all data analysis studies of $(\ell, m) = (2,2)$ ringdown signals neglect
the negative-frequency QNMs as they are not expected to be excited with appreciable power~\cite{abbott2021tests,Berti:2005ys,isi2021analyzing,capano2021observation}. Such expectations, however, typically draw from the comparable mass ratio regime where the remnant black hole's spin has a positive projection onto the orbital angular
momentum at the plunge~\cite{zertuche2022high}. For systems with large mass ratios (where black hole perturbation theory becomes increasingly applicable), we have $\chi_1 \approx \chi_f$, whence the remnant black hole will counter-rotate relative to the orbit whenever $\chi_1 < 0$. We will demonstrate that both positive and negative-frequency QNM modes can be excited in the context of point-particle perturbation theory (confirming previous work of Refs.~\cite{lim2019exciting, Hughes:2019zmt,taracchini2014small,Lim:2022veo}) whenever $\chi_1$ is sufficiently negative. We then show evidence for the same qualitative behavior in a recent large mass ratio numerical relativity simulation. 

Following Ref.~\cite{Varma:2018mmi}, we compute the Fourier transform of $h_{22}$ (computed in perturbation theory) in the ringdown stage of the waveform for different values of mass ratio and spin. Before computing the Fourier transform, we first drop all data before $t = 15 m_1$, where $t = 0$ corresponds to the peak of the waveform amplitude according to Eq.~\eqref{eq:amplitude}. Then, we taper the data between $t = 15m_1$ and $t = 35m_1$ and the last $10m_1$ of the time series using a Planck window.

\begin{figure*}[htb]
	\centering
	\subfigure[]{\label{fig:QNM_retrograde_EMRI_vsQ}
		\includegraphics[scale=0.49]{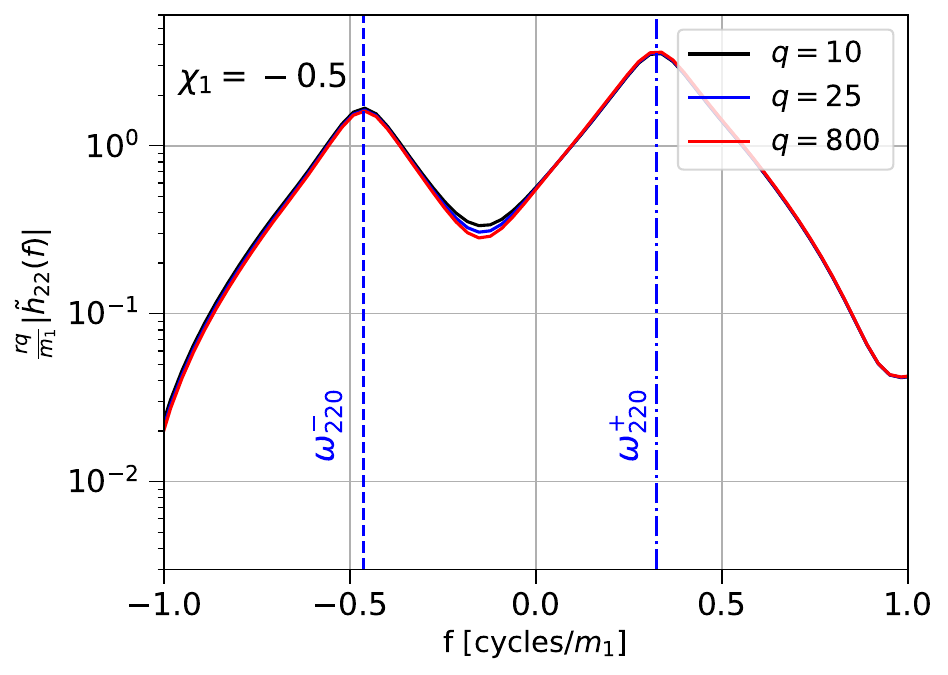}}
	\subfigure[]{\label{fig:ringdown_QNM_EMRI}
		\includegraphics[scale=0.49]{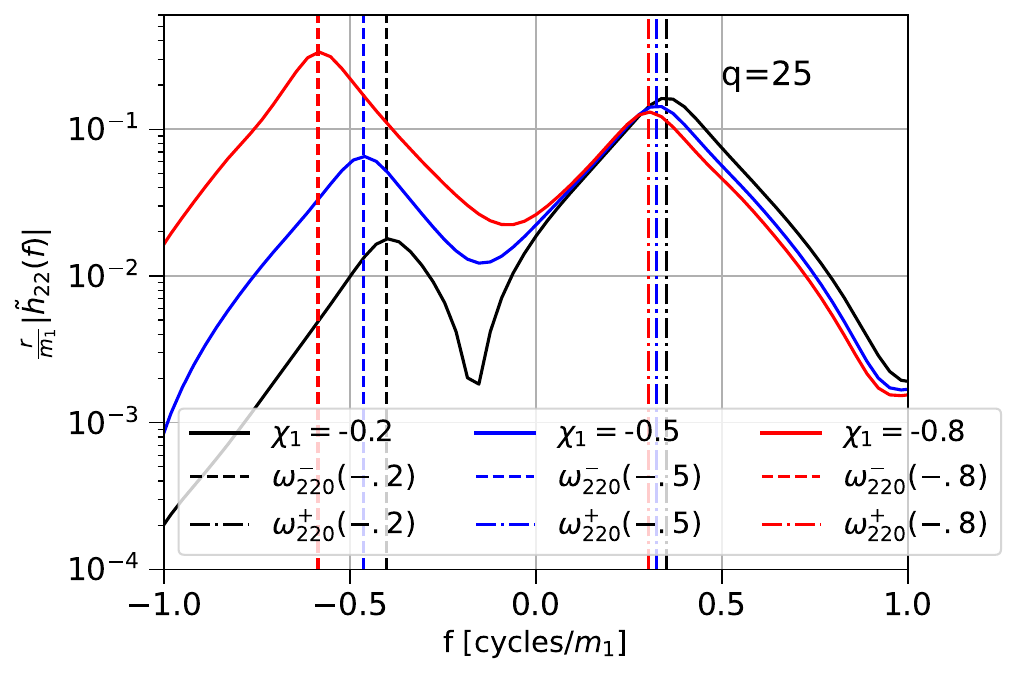}}
	\caption{\label{fig:retrograde_EMRI} The absolute value of the Fourier transform of a ringdown signal, $h_{22}$, computed using point-particle perturbation theory. In both figures, the vertical line depicts the real part of the $(\ell, m,n) = (2,2,0)$ negative-frequency (dashed; $\omega_{220}^{-}$) and positive-frequency (dash-dot; $\omega_{220}^{+}$) QNM for a Kerr BH with a spin of $\chi_1$.
 {\bf Left}: Systems with a fixed spin of $\chi_1 = -0.5$ while varying the mass ratio from $q=10$ to $q=800$. Due to the amplitude's large-$q$ scaling, $h \propto q^{-1}$, we scale the amplitudes by $q$ in addition to the usual $r/m_1$ scaling. As expected and explained in the text, the signal shows no dependence on the mass ratio. We find that the positive-frequency QNM is excited for these systems with an amplitude nearly comparable to the negative-frequency QNM. 
 {\bf Right}: Systems with a fixed mass ratio of $q = 25$ while varying the spin from $\chi_1=-0.2$ to $\chi_1=-0.8$. The amplitude associated with the negative-frequency QNM increases as the spin becomes increasingly negative, and by about $\chi_1 \approx -0.6$, the negative-frequency QNM dominates over the positive-frequency one, resulting in the phase reversal seen in Fig.~\ref{fig:ringdown}. }
\end{figure*}

Figure~\ref{fig:QNM_retrograde_EMRI_vsQ} shows the absolute value of these Fourier transformed signals for systems with a fixed spin of $\chi_1 = -0.5$ while varying the mass ratio from $q=10$ to $q=800$. We find both positive-frequency and negative-frequency QNMs to be excited with comparable amplitude~\footnote{The QNM frequencies are computed with the {\tt pyKerr} package~\cite{collin_capano_2023_10056495}, which interpolates the tables of Berti et al.~\cite{Berti:2005ys} that have been compiled using the analytic methods of Leaver~\cite{leaver1985analytic}. We have also confirmed the {\tt pyKerr} computation by comparing it to the  {\tt qnm} package~\cite{Stein:2019mop}. Note that  {\tt pyKerr} and  {\tt qnm} have different conventions that must be accounted for when evaluating the $m<0$ QNM values.}. Furthermore, as expected, the signal shows no dependence on the mass ratio. This can be understood by noting that in point-particle black hole perturbation theory, (i) the background spacetime is fixed, (ii) the remnant black hole is exactly $(m_1, \chi_1)$, and (iii) the secondary black hole's plunge trajectory is a geodesic, that is it is independent of mass ratio. 

Figure~\ref{fig:ringdown_QNM_EMRI} shows the absolute value of these Fourier transformed signals for systems with a fixed mass ratio of $q = 25$ while varying the spin from $\chi_1=-0.2$ to $\chi_1=-0.8$. In this case, the QNM excitations have a spin dependence, and we find that the negative-frequency QNM's power increases as the spin becomes increasingly negative. By about $\chi_1 \approx -0.6$, the negative-frequency mode's amplitude is larger than the positive-frequency mode, resulting in the phase reversal seen in Fig.~\ref{fig:ringdown}.

\begin{figure}
    \centering
    \includegraphics[width=8cm]{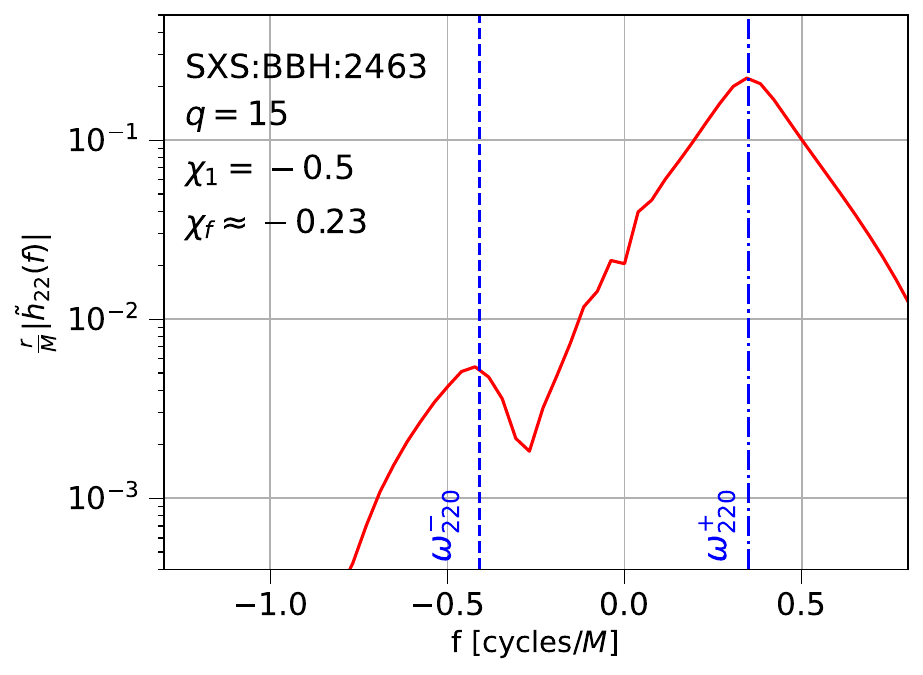}
    \caption{The absolute value of the Fourier transform of a ringdown signal, $h_{22}$, computed using the numerical relativity code SpEC. This particular BBH system has one of the most negative remnant spins, $\chi_f \approx -0.23$, in the public SXS catalog, making it well suited for looking for the appearance of both positive-frequency and negative-frequency QNMs. We find two peaks in the Fourier spectrum at frequency values expected for a ringdown signal described by two QNMs.  The vertical line depicts the real part of the $(\ell, m,n) = (2,2,0)$ negative-frequency (dashed; $\omega_{220}^{-}$) and positive-frequency  (dash-dot; $\omega_{220}^{+}$) QNM for a Kerr BH with a mass of $M_f$ and spin of $\chi_f$.  Comparing to the point-particle perturbation theory result of Fig.~\ref{fig:retrograde_EMRI}, we see that the NR ringdown data is qualitatively most similar to the $\chi_1=-0.2$ case.
    }
    \label{fig:QNM_retrograde_SXS}
\end{figure}

At what mass ratio can the ringdown system predicted by perturbation theory describe realistic systems? To investigate this further, we use a recent high mass ratio numerical relativity simulation~\cite{yoo2022targeted} with $q=15$ and $\chi_1=-0.5$. For this system, the remnant spin was measured to be $\chi_f \approx -0.23$, so we might expect the ringdown dynamics to resemble a perturbation theory waveform for a system with $\chi_1 \approx -0.23$. Fig.~\ref{fig:QNM_retrograde_SXS} largely confirms this expectation and provides evidence for a non-negligible negative-frequency QNM. It is worth noting that, as expected, the NR simulation data does not resemble the corresponding $\chi_1 = -0.5$ ppBHPT waveform shown in Fig.~\ref{fig:ringdown_QNM_EMRI}.

Due to these known limitations in the ppBHPT computation, our model \model will have systematic errors in the ringdown portion of the waveform for comparable to moderate mass ratios. More specifically, we expect modeling errors in the ringdown signal to be related to the difference between the initial and remnant spin, $\Delta(\chi_1, q) = |\chi_1 - \chi_f(\chi_1, q) |$. Recently built remnant model for large mass ratio systems~\cite{boschini2023extending} can be used to compute $\Delta$, although it is currently unclear how to bound the ringdown modeling error for our NR-calibrated model \model by $\Delta$.

\section{Resolvability of the waveform's peak}
\label{app:peaktime}

Before building our model, a simulation-dependent time shift is applied to all $476$ ppBHPT training waveforms so that the maximum of the total amplitude, defined by Eq.~\eqref{eq:amplitude}, occurs at $t=0$. The accuracy with which we can align a set of waveforms depends on our ability to accurately compute $\tau_\mathrm{peak}$, and our procedure for this is summarized in Sec.~\ref{subsec:dataAlignement}. The surrogate model's accuracy, in turn, depends crucially on this alignment. As demonstrated in Figure 15 of Ref.~\cite{Field:2013cfa}, for example, the surrogate model's error is proportional to the typical uncertainty in the peak $\Delta \tau_\mathrm{peak}$. If $\Delta \tau_\mathrm{peak}$ becomes large, it can become the dominant source of model error. 

One way to estimate $\Delta \tau_\mathrm{peak}$ is to compute the peak of the waveform after interpolating the modes onto a common time grid with a cubic spline (cf. Sec.~\ref{subsec:dataAlignement}). Absent numerical error, each waveform's peak should be exactly $t_\mathrm{peak}=0$ and
so we can easily estimate the error.

\begin{figure}
    \centering
    \includegraphics[width=8cm]{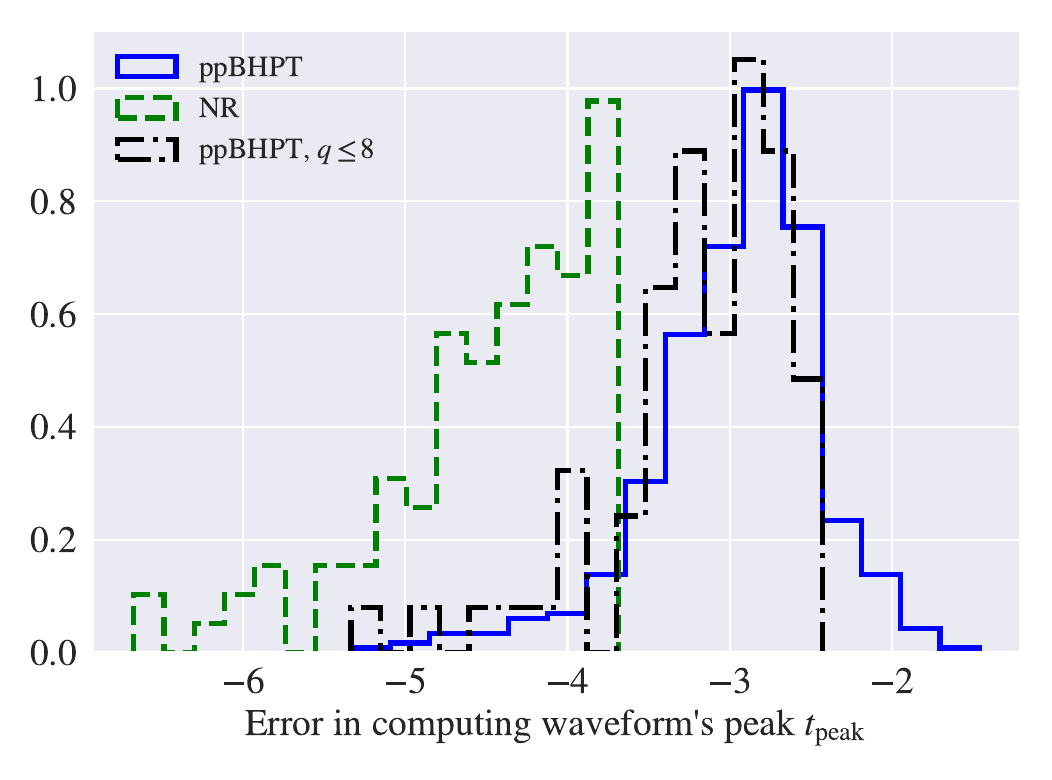}
    \caption{Estimated error in measuring the time at which the waveform's amplitude reaches a maximum. Errors are computed for each waveform in the training set used to build the numerical relativity surrogate model \texttt{NRHybSur3dq8}~\cite{Varma:2018mmi} (labeled ``NR''), the ppBHPT-based surrogate model \model considered in this paper (labeled ``ppBHPT''), and \model restricted to the region of the parameter space also covered by \texttt{NRHybSur3dq8}. The increased difficulty in revolving the peak of a ppBHPT waveform can show up as an overall modeling error.}
    \label{fig:tpeak}
\end{figure}

Fig.~\ref{fig:tpeak} compares the error in computing the peak time for each of the $476$ ppBHPT training waveforms with the error in computing the peak time for each of the 91 simulations used to train the aligned-spin NR surrogate model \texttt{NRHybSur3dq8}~\cite{Varma:2018mmi}. Despite using identical codes for waveform alignment, we find higher accuracies in resolving the peak time in the NR training data. This is likely due to differences in the training data. First, the ppBHPT simulations go to much higher mass ratios where the peak tends to flatten out. Restricting the ppBHPT simulations to the same regime as the NR surrogate model ($q \leq 8$), the errors become a bit more comparable. A second difference is that the NR waveforms are specified on an adaptive time grid with high temporal resolution near the merger, while the ppBHPT training waveforms use a uniform time grid of spacing $\Delta t = 0.08 m_1$. While there are potentially many other differences between the two waveform sets, we believe these two differences are primarily responsible for the observed differences. Future high-accuracy models based on ppBHPT may need additional temporal resolution near the merger to reduce this potential source of modeling error. 

\section*{References}

\bibliography{References}

\end{document}